\documentclass[manuscript]{aastex}
\usepackage{amsmath,amsfonts,amssymb,amscd,amsthm,amsbsy, color}
\usepackage{graphicx}
\usepackage{courier}
\usepackage{fancybox}
\usepackage{multimedia}
\usepackage{graphicx,epstopdf}
\usepackage{appendix}
\usepackage{wasysym}

\newcommand\beq[1]{ \begin{equation}\label{#1} }
\newcommand{\eeq}{ \end{equation} }

\newcommand\beqa[1]{ \begin{eqnarray} \label{#1}}

\newcommand{\eeqa}{ \end{eqnarray} }
\newcommand{\beqano}{ \begin{eqnarray*} }
\newcommand{\eeqano}{ \end{eqnarray*} }
\newcommand\equ[1]{{ \rm (\ref{#1})}}
\newcommand\RI[1]{{\textcolor{black} {#1}}}

\slugcomment{Draft}

\shorttitle{Secular motion of charged dust grains in the solar system}
\shortauthors{Lhotka et al.}

\begin{document}

\title{Charged dust grain dynamics subject to solar wind, Poynting-Robertson drag,
and the interplanetary magnetic field}

\author{
Christoph Lhotka\altaffilmark{1} 
and
Philippe Bourdin
and
Yasuhito Narita
}
\affil{
Space Research Institute, Austrian Academy of Sciences, Schmiedlstrasse 6,
A-8042 Graz}
\email{christoph.lhotka@oeaw.ac.at}
\email{philippe.bourdin@oeaw.ac.at}
\email{yasuhito.narita@oeaw.ac.at}
\altaffiltext{1}{Corresponding author.}

\date{Received: date / Accepted: date}

\begin{abstract}
We investigate the combined effect of solar wind, Poynting-Robertson drag, and
the frozen-in interplanetary magnetic field on the motion of charged dust
grains in our solar system. For this reason we derive a secular theory of
motion by the means of averaging method and validate it with numerical
simulations of the un-averaged equations of motions. The theory predicts that
the secular motion of charged particles is mainly affected by the z-component
of the solar magnetic axis, or the normal component of the interplanetary
magnetic field. The normal component of the interplanetary magnetic field leads
to an increase or decrease of semi-major axis depending on its functional form
and sign of charge of the dust grain.  It is generally accepted that the
combined effects of solar wind and photon absorption and re-emmision
(Poynting-Robertson drag) lead to a decrease in semi-major axis on secular time
scales. On the contrary, we demonstrate that the interplanetary magnetic field
may counteract these drag forces under certain circumstances. We derive a
simple relation between the parameters of the magnetic field, the physical
properties of the dust grain as well as the shape and orientation of the
orbital ellipse of the particle, which is a necessary conditions for the
stabilization in semi-major axis.
\end{abstract}

\keywords{Interplanetary magnetic field, solar wind drag, Poynting-Robertson drag,
dust grains, celestial mechanics}


\section{Introduction}

Micrometer-sized particles in the solar system originate from asteroid collisions
and cometary activities, and can be found at various places in interplanetary
space.  The dynamics of uncharged micrometer-sized particles is influenced by
different effects.  First, the gravitational force attracts particles toward
the Sun and the planets.  Second, the solar radiation pressure pushes particles
away from the Sun.  Third, the combined solar wind and Poynting-Robertson
effect brakes the particle motion due to a momentum transfer.  These forces
compete against one another such that some particles may fall onto the Sun, some
leave the solar system, and some stay temporarily captured in the solar system
\citep[see, e.g.][]{ManEtAl2006,ManEtAl2014}. \\

\RI{Dust grains get charged by collecting and emitting charged particles. As a
result the net charge and electrostatic potential of the dust grains change
with time too.  These changes finally end if a charge equilibrium has been reached.
The main charging mechanisms in the solar system are as follows: i) impacts of
electrons and ions directly transfer their charge to the grain; ii)
photo-ejection of electrons by ultra-violet radiation of the Sun, as well as
iii) recombination with free electrons from the dust grain environment. In the solar
system the grains typically obtain a positive charge due to the dominance of
ii) that corresponds to values of the electric potential of the order of 1-10
Volts, i.e. 5 Volts for grains around 1 micron in diameter.}   \\

Here, we conduct an analytical and numerical study on the impact of the normal
magnetic field component, with respect to the equator of the Sun, to the
orbital dynamics of micrometer-sized particles in the interplanetary magnetic
field. Usually, the influence of the normal component of the interplanetary
field on the micrometer-sized particle dynamics is neglected.  Naively speaking,
one may anticipate that the magnetic field imposes a change in particle orbits
(e.g., the inclination, the eccentricity, or the semi-major axis) through
gyro-motion around the magnetic field, the adiabatic change against the
inhomogeneous magnetic field, and drift motions in a gravitational field.  We start
with the equation of motion for the electrically charged, micrometer-sized
particles and investigate the time evolution of the orbits, as well as an
equilibrium state imposed by the magnetic field, both incorporating an axi-symmetric
spiral shape of the planar interplanetary magnetic field \citep{Parker1958,weber1967}.
We find that the normal component of the magnetic field with respect to the
equatorial plane of the Sun, i.e. the orbital plane of the particle plays a
crucial role. \RI{The likely magnitude of the normal magnetic field component
can be estimated on the basis of the fast pole to pole transit of the Ulysses 
spacecraft at solar minimum activity. We make use of the analysis in \citet{For1996, For2002} 
of the spacecraft data to find representative values of the magnetic field
components. A 3-dimensional model of the heliospheric magnetic field has also been 
used in \citet{Zur1997}, where the authors find that the normal component scales 
with the inverse distance between the particle and the magnetic source.} \\

The secular orbital evolution of dust grains due to solar wind and
Poynting-Robertson drag alone has been investigated, e.g. in \citet{Kla2013}.
An important contribution to the research of motion of charged dust particles
in interplanetary fields can be found in \citet{MorGru1979b, MorGru1979a}.
Here, the authors provide a detailed analysis of various systematic effects
caused by electromagnetic forces on the orbital parameters of charged dust
grains, i.e.  they show that small stochastic variations induced by these
forces are unimportant for particles of sufficient mass.  The interplay between
drag and Lorentz forces has been investigated in \citet{MukGie1984}, where the
authors find that the Lorentz force introduces a significant effect on the
orbital inclination $i$, while the effect on the variation in \RI{semi-major axis} $a$ is
negligible. Stochastic diffusion of interplanetary dust grains orbiting under
Poynting-Robertson drag force and within the interplanetary field has been
discussed in \citet{WalHas1985}.  In \citet{FahEtAl1995} the authors find that
particle distributions depend on inclination and distance from the Sun in the
case of asymmetric solar winds.  The drift in inclination $i$ due to
electromagnetic force has been studied in detail in \cite{FahEtAl1981}.  This
work predicts that dust can be expected to be concentrated close to the
magnetic equator. The dynamics of dust in the vicinity of the Sun has been
treated in \citet{KriEtAl1998}, where the authors find that the orientation of
the orbital planes of the particles is dictated by electromagnetic forces.
Typical dynamical evolution of charged dust particles has also been treated in
\citet{KriManKim1998}. The authors find that the radial motion of particles are
relatively insensitive to the electromagnetic force, while orbital planes are
affected depending on the size and chemical composition.  The interactions of
dust grains with coronal mass ejections and solar cycle variations are analyzed
in \citet{RagKah2003}. Numerical simulations of particle orbits subject to
Lorentz force, solar wind, and Poynting-Robertson drag can be found in
\citet{KocKla2004,Koc2006}. In the latter, the authors mainly focus on the
temperature-dependent dielectric functions of carbonaceous or silicate
particles, but also provide a numerical study of the long-term dynamics of
micrometer-sized particles with changing optical properties. In
\citet{ManMurCze2007} the authors show that nanometer-sized particles can stay
in bound orbits and, aside from the Lorentz force, the plasma and the photon
Poynting-Robertson effect determine their spatial distribution.

Closest related with our study is \cite{Con1979}, where the mean square change
in the orbital elements due to electromagnetic interactions is compared with
the net Poynting-Robertson effect. The author states: {\it ''Lorentz scattering
can maintain significant numbers of micron and submicron particles against loss
from the solar system due to Poynting-Robertson drag.''}. While, the author
further wrote {\it ''...there is no obvious way to calculate the magnitude of these 
secular changes...''} (of the orbital elements), the author already estimated
that {\it ''...there may well be secular changes in the orbital elements due to 
Lorentz force.''}. \\

With our study we would like to quantify these statements. We describe the
secular evolution of the orbital elements due to Lorentz force by means of
\RI{Equations} \equ{secgauss},\equ{secgauss2} in Section 2.3., and we provide a
charge over mass ratio to balance the Lorentz force with solar wind and
Poynting-Robertson drag force by means of \RI{Equation} \equ{formula} in Section 2.5.

\section{The dynamical model}
\label{model}

We investigate the dynamics of micro-meter sized, spherical particles of
radius $R$, density $\rho$, and mass $m$ orbiting in our solar system.

\subsection{Set-up of notation and Cartesian framework}

Let ${\mathbf e}_x=(1,0,0)$, ${\mathbf e}_y=(0,1,0)$, ${\mathbf e}_z=(0,0,1)$
be Cartesian unit vectors in a heliocentric coordinate system. We denote by
$\mathbf r$, $\mathbf v$ the Cartesian position and velocity of the dust grain
with scalar distance $r=\|{\mathbf r}\|$ from the Sun. Moreover, $\mu\equiv GM$ is
the heliocentric gravitational constant, $S$ is the solar energy flux at
distance $r$, $A$ is the cross-sectional area of the
grain, $Q$ is the spectrally averaged dimensionless efficiency factor of the
radiation pressure, and $c$ is the speed of light. We introduce the ratio of solar
radiation pressure over solar gravititational attraction:

\beq{beta}
\beta=\frac{SA Q}{c}/\frac{GMm}{r^2} \ .
\eeq

Since $S\propto1/r^2$ we notice that $\beta$ is a dimensionless parameter
without radial dependence. Let ${\mathbf u}_{sw}$ be the velocity vector of the
solar wind with magnitude $u_{sw}$, and $\eta$ be the dimensionless solar wind
drag efficiency factor (the ratio of solar wind over Poynting-Robertson drag).
Furthermore, ${\mathbf r}_1$ is the Cartesian position of an additional planet
of mass $m_1$ in the heliocentric reference frame.  Moreover, we denote by
$B_0$ the magnetic field strength at the reference distance $r_0$ of the
magnetic field $\mathbf B$ that originates from the Sun. The equation of motion 
of the particle of charge $q=4\pi\varepsilon_0UR$  is given by:

\beqa{EOM}
\frac{d{\mathbf v}}{dt}&=&-\frac{(1-\beta)\mu}{r^2}{\mathbf e}_R
-\frac{\beta\mu}{r^2}\left(1+\frac{\eta}{Q}\right)
\left(\frac{({\mathbf v}\cdot {\mathbf e}_R){\mathbf e}_R+\mathbf v}{c}\right) \\
&-&{ G}m_1\left(
\frac{{\mathbf r}_1}{r_1^3}+
\frac{{\mathbf r}-{\mathbf r}_1}
{\|{\mathbf r}-{\mathbf r}_1 \|^3}\right)
+\frac{q}{m}
\left(\mathbf v-\mathbf u_{sw}\right)\times\mathbf{B} \ , \nonumber
\eeqa   

where $\varepsilon_0$ is the permitivity of vacuum, and $U$ denotes the particle's surface
potential. Let $\boldsymbol\omega=(\omega_1, \omega_2, \omega_3)$ be the
direction of the magnetic axis of the Sun given in the heliocentric frame of
reference. We denote by $\mathbf e_R={\mathbf r}/r$, ${\mathbf
e}_T={\boldsymbol \omega}\times{\mathbf e}_R$, and ${\boldsymbol
e}_N=\boldsymbol\omega$ the radial, tangential, and normal unit vectors in a
body fixed reference frame attached to the Sun.
The magnetic field can be decomposed in terms of the radial $B_R$, 
tangential $B_T$, and normal component $B_N$ as follows:

\beq{B}
{\mathbf B} = 
B_R{\mathbf e}_R + 
B_T{\mathbf e}_T + 
B_N{\mathbf e}_N \ .
\eeq

\noindent In this framework, a radially expanding and uniform solar wind is given by:
\beq{SW}
{\mathbf u}_{sw} = u_{sw}{\mathbf e}_R \ .
\eeq
 
Equation \equ{EOM} reduces for $\beta=0$, $m_1=0$, and $q/m=0$ to the
integrable two-body problem.  For $\beta\neq0$ the first term corresponds to
the solar radiation pressure of the two-body problem with reduced central mass
$(1-\beta)$.  The second term is the sum of the drag terms due to solar wind
friction and the Poynting-Robertson effect. The third term is the gravitational
perturbation due to the additional planet. The last term is the
acceleration that a charged grain experiences due to the presence of a ''frozen-in'' magnetic
field, and in absence of a background electric current resistivity (${\mathbf
E}=-{\mathbf u}_{sw}\times{\mathbf B}$). ''Frozen-in'' refers to the magnetic
and thermal plasma pressure, where in the solar wind the magnetic field is frozen-in
to the plasma bulk motion.

\subsection{Gauss' form of the equations of motions}

We denote the orbital elements of the particle by the semi-major axis $a$, the
eccentricity $e$, the inclination $i$, the argument of perihelion $\omega$, the
longitude of the ascending node $\Omega$, true anomaly $f$, and the mean
anomaly $M$. The norm of the orbital angular momentum is then given by
$h=\sqrt{1-e^2}\sqrt{\mu a}$. Kepler's 3rd law is $\mu=n^2a^3$, where $n$ is
the mean motion of the dust grain. In this setting, Gauss' form of perturbed
equations of  motion are given by \citep[c.f.][]{Fitzpatrick}:

\beqa{gauss}
\frac{da}{dt} &=& 
\frac{2ah}{\mu\left(1-e^2\right)}
\left[e\sin f F_R + \left(1 + e\cos f\right) F_T\right] \ , \\
\frac{de}{dt} &=& 
\frac{h}{\mu}
\left[\sin f F_R + \left(\cos f + \cos E\right) F_T\right] \ ,  \nonumber  \\
\frac{di}{dt} &=& 
\frac{\cos\left(\omega+f\right)r}{h} F_N , \ \nonumber \\
\frac{d\omega}{dt} &=&
-\frac{h}{\mu}\frac{1}{e}
\left[\cos f F_R-\left(\frac{2+e\cos f}{1+e cos f}\right)\sin f F_T\right]
-\frac{\cos i\sin\left(\omega+f\right)r F_N}{h \sin i} \ , \nonumber \\
\frac{d\Omega}{dt} &=&
\frac{\sin\left(\omega+f\right)r}{h \sin i} F_N \ , \nonumber \\
\frac{dM}{dt} &=& n +
\frac{h}{\mu}\frac{\sqrt{1-e^2}}{e}
\left[\left(\cos f-\frac{2e}{1-e^2}\frac{r}{a}\right) F_R
-\left(1+\frac{1}{1-e^2}\frac{r}{a}\right)\sin f F_T\right] \nonumber \ .
\eeqa

Here, $F_R$, $F_T$, $F_N$ are the radial, tangential, and normal components of
the perturbing force ${F}=F_R {\mathbf e'}_R + F_T {\mathbf e'}_T + F_N
{\mathbf e'}_N$ given in the orbital frame of the particle: ${\mathbf
e'}_R=(\cos f,\sin f,0)$, ${\mathbf e'}_T=(-\sin f, \cos f, 0)$, and ${\mathbf
e'}_N={\mathbf e'}_R\times{\mathbf e'}_T$. The relation between orbital and
heliocentric reference frame is provided by the rotation matrix\footnote{
The rotation matrices are defined as follows:
\beqano
{\mathfrak R}_1(\varphi)=\left(
\begin{array}{ccc}
 1 & 0 & 0 \\
 0 & \cos\varphi & -\sin\varphi \\
 0 & \sin\varphi & \cos\varphi \\
\end{array}
\right) \ , \quad
{\mathfrak R}_3(\varphi)=\left(
\begin{array}{ccc}
 \cos\varphi & -\sin\varphi & 0 \\
 \sin\varphi & \cos\varphi & 0 \\
 0 & 0 & 1 \\
\end{array}
\right) \ .
\eeqano}

\beq{dy1}
{\mathfrak
R}={\mathfrak R}_3(\Omega)\cdot{\mathfrak R}_1(i)\cdot{\mathfrak R}_3(\omega) \ .
\eeq

For a generic force given in the heliocentric reference frame in terms of 
${\mathbf F}=F_x {\mathbf e}_x + F_y {\mathbf e}_y + F_z {\mathbf e}_z$ the
following relations hold true:

\beqa{dy2}
F_R = \left(F_x,F_y,F_z\right) \cdot {\mathfrak R} \cdot {\mathbf e'}_R \ , \ 
F_T = \left(F_x,F_y,F_z\right) \cdot {\mathfrak R} \cdot {\mathbf e'}_T \ , \ 
F_N = \left(F_x,F_y,F_z\right) \cdot {\mathfrak R} \cdot {\mathbf e'}_N \ .
\eeqa

Thus, if we identify ${\mathbf F}$ with the perturbing parts of the Kepler
problem in \RI{Equation} \equ{EOM} (i.e. without the term that defines the unperturbed two
body problem) then it is true that \RI{Equations} \equ{gauss} are equivalent with the
equations of motion defined by \RI{Equation} \equ{EOM}. Using well known formulae for Taylor
series expansions in the two-body problem \citep[c.f.][]{mybook} the right
hand sides of \RI{Equation} \equ{gauss} can be written in terms of orbital elements of the
dust particle and the perturbing planet, only. For $\beta=0$, $m_1=0$, and
$q/m=0$ the system \RI{of Equations} \equ{gauss} reduces to the single integrable 
equation of motion $dM/dt=n$.  \\

\subsection{The magnetized two body problem}
\label{M2BP}

\cite{LhoCel2015} have treated the influence of solar radiation pressure, solar
wind, and Poynting-Robertson drag force. In this Section, we are mainly
interested in additional effects of the Lorentz-force. Let the components
of a generic interplanetary magnetic field in \RI{Equation} \equ{B} be the product 
of constant, radially dependent, and time dependent terms:

\beq{Bex}
B_R=\RI{B_{R0}}\left(\frac{r_0}{r}\right)^2b_R\left(t\right) \ , \quad
B_T=\RI{B_{T0}}\left(\frac{r_0}{r}\right)b_T\left(t\right) \ , \quad
B_N=\RI{B_{N0}}\left(\frac{r_0}{r}\right)^\kappa b_N\left(t\right) \ .
\eeq

Here, \RI{$B_{R0}$, $B_{T0}$, $B_{N0}$ are the components of} the average magnetic field 
background strength at reference
distance $r_0$, and $b_R$, $b_T$, $b_N$ are periodic functions in time, that
are introduced to mimic the solar cycle. The radial dependencies $1/r^2$ in
$B_R$ and $1/r$ in $B_T$ resemble those of the classical Parker spiral
\citep[c.f.][]{Parker1958, GruEtAl1994, Mey2007}. Assuming a radial magnetic
field of the source and a purely radial expansion of the solar wind it is
possible to set $B_N=0$ and we recover the time dependent Parker spiral
\citep[c.f.][]{Koc2006} with:

\beq{dy3}
b_R\left(t\right) = \cos\left(2\pi t/T+\varphi_0\right) \ , \quad 
b_T\left(t\right) = \cos\left(\vartheta\right)\cos\left(2\pi t/T+\varphi_0\right) \ ,
\eeq

where $\varphi_0$ is the magnetic phase angle and $\vartheta$ is the altitude
from the solar equatorial plane. $T$ is the period of the solar magnetic cycle
equal to about $22$ years. Here, we now add the effect of a non-zero normal
component of the magnetic field $B_N\neq0$ on the particle dynamics.
\RI{According to \citet{Zur1997} the normal component scales with $1/r$, i.e.
$\kappa=1$ in Equation \equ{Bex}.  To understand the role of the exponent in
$1/r$ on the secular motions we also include $\kappa=2,3$ in our study. We
notice that $B_N$ is not necessarily consistent with $\nabla\cdot \mathbf B=0$
unless we add small extra dependencies that could lead to a consistent value
for $B_N$. However, we argue that these additional contributions are small on
average and do not alter our results on secular time scales, i.e. on the basis
of a mean interplanetary magnetic field.} Let $B_R'=B_{R0}\cdot b_R(t)$,
$B_T'=B_{T0}\cdot b_T(t)$, and $B_N'=B_{N0}\cdot b_N(t)$. If we plug in \RI{Equation} \equ{Bex}
in \RI{Equation} \equ{B}, the right hand sides of the \RI{first three of Equations}
\equ{gauss} become of the form:

\beqa{LFgauss}
\frac{da}{dt} &=& 
\frac{q}{m}\frac{u_{sw}}{n}
\left(
\left[\frac{r_0}{a}\right]B_T'\left[1\right] +
\left[\frac{r_0}{a}\right]^\kappa
\left(c_{\mathbf 0}^{(2,3)}\omega_3+\left[2\right]\right)B_N'
\right)  \\
\frac{de}{dt} &=& 
\frac{q}{m}
\bigg(
\left\{
\frac{u_{sw}}{n}
\left[\frac{r_0}{a}\right]^2
\left[3\right] +
\left[\frac{r_0}{a}\right]
\left[4\right]
\right\}B_T' \nonumber \\
&+&
\left\{
\frac{u_{sw}}{na}
\left[\frac{r_0}{a}\right]^{\kappa}
\left(
c_{\mathbf 0}^{(5,3)}\omega_3+
\left[5\right]
\right)+
\left[\frac{r_0}{a}\right]^\kappa
\left[6\right]
\right\}B_N'
\bigg)
\nonumber \\
\frac{di}{dt} &=& 
\frac{q}{m}\bigg(
\left[\frac{r_0}{a}\right]^2
\left[A\right]
B_R'+
\left\{
\frac{u_{sw}}{n}
\left[\frac{r_0}{a}\right]^2
\left[7\right] +
\left[\frac{r_0}{a}\right]
\left[8\right]
\right\}B_T' \nonumber \\
&+&
\left\{
\frac{u_{sw}}{na}
\left[\frac{r_0}{a}\right]^{\kappa}
\left(
c_{\mathbf 0}^{(9,3)}\omega_3 +
\left[9\right]
\right) +
\left[\frac{r_0}{a}\right]^\kappa
\left[10\right]\right\}B_N'
\bigg) \ .
\nonumber
\eeqa

Here, each term $[\#]$ in \RI{Equation} \equ{LFgauss} takes the structural form:

\beq{dy4}
[\#]=\sum_{\ell=1,2,3}
\omega_\ell
\bigg\{
\sum_{{\mathbf k}\in{\mathbb Z}^3}
\begin{array}{c}
c_{\mathbf k}^{(\#,\ell)} \\
s_{\mathbf k}^{(\#,\ell)} \\
\end{array}
\big(e,i\big)
\begin{array}{c}
\cos{} \\
\sin{} \\
\end{array}
\left(k_1 M+k_2 \omega+k_3\Omega\right)
\bigg\} \ ,
\eeq

while the term $\left[A\right]$ proportional to the only radial contribution $B_R$
in \RI{Equation} \equ{LFgauss} becomes:

\beq{dy6}
[A]=\sum_{{\mathbf k}\in{\mathbb Z}^3}
\begin{array}{c}
c_{\mathbf k}^{(A)} \\
s_{\mathbf k}^{(A)} \\
\end{array}
\big(e,i\big)
\begin{array}{c}
\cos{} \\
\sin{} \\
\end{array}
\left(k_1 M+k_2 \omega+k_3\Omega\right) \ .
\eeq

Second order Taylor series expansions of the coefficients $c_{\mathbf
k}$, $s_{\mathbf k}$, that are valid in small eccentricity $e$, are provided
in Appendix A.  We notice, that the radial contribution $B_R$ appears only in the
equation for $di/dt$ in \RI{Equation} \equ{LFgauss}. Moreover, the radial component of the
magnetic field is independent of the choice of $\boldsymbol\omega$, as
expected. In addition, while for a vanishing solar wind speed $da/dt=0$, the
components $de/dt$, $di/dt$ do not vanish. \\

The terms $[\#]$ and $[A]$ are periodic functions of zero average and
wave number $\mathbf k$. Using  averaging theory, we may
neglect these terms, and investigate the long term dynamics by
setting $[\#]=0$, $[A]=0$ in \RI{Equation} \equ{LFgauss}. We obtain the secular system:

\beqa{secgauss}
\frac{da}{dt} &=& \omega_3B_N'\left(\frac{r_0}{a}\right)^\kappa\frac{q}{m}\frac{u_{sw}}{n}
g_{\kappa,a}\left(e\right)\cos i \ ,
\nonumber \\
\frac{de}{dt} &=& \omega_3B_N'\left(\frac{r_0}{a}\right)^\kappa
\frac{q}{m}\frac{u_{sw}}{na}
g_{\kappa,e}\left(e\right)\cos i \ ,
\nonumber \\
\frac{di}{dt} &=&
\omega_3B_N'\left(\frac{r_0}{a}\right)^\kappa\frac{q}{m}\frac{u_{sw}}{na}
g_{\kappa,i}\left(e\right)\sin i \ ,
\eeqa

where $g_{\kappa,a}$, $g_{\kappa,e}$, $g_{\kappa,i}$ are functions of
eccentricity $e$, originating from the coefficients $c_{\mathbf 0}^{(2,3)}$,
$c_{\mathbf 0}^{(5,3)}$, and $c_{\mathbf 0}^{(9,3)}$, respectively. A Taylor
series expansion of order $4$ in small eccentricity $e$ can be found in
Appendix A. \\

We immediately recognize, that only the normal component $B_N$ of \RI{Equation}
\equ{Bex} can lead to a secular motion, i.e. drift in the semi-major axis, the
eccentricity, and inclination on secular time scales. An order of magnitude
comparison of the right hand sides of \RI{Equations} \equ{secgauss} shows that $de/da=1/a$,
and $di/dt\to0$ for small inclination $i\to0$.  \\

{\it What about the qualitative behaviour of the angle-like Kepler elements
$\omega$, $\Omega$, $M$?} It turns out, that the right hand side of
$d\omega/dt$, $d\Omega/dt$ in \RI{Equation} \equ{gauss} are affected in a similar way as
$di/dt$, i.e. by radial, tangential, and normal components of the magnetic
field, while the right hand side of $dM/dt$ is not affected by the radial
component $B_R$ (comparable to the case $da/dt$ described above). The same approach as
for the action-like Kepler elements $a$, $e$, $i$, leads us to a secular
system of the angle-like variables $\omega$, $\Omega$ and $M$:

\beqa{secgauss2}
\frac{d\omega}{dt} &=&
\omega_3B_N'\left(\frac{r_0}{a}\right)^\kappa\frac{q}{m}
g_{\kappa,\omega}\left(e\right)\cos i \ ,\nonumber \\
\frac{d\Omega}{dt} &=&
\omega_3B_N'\left(\frac{r_0}{a}\right)^\kappa\frac{q}{m}
g_{\kappa,\Omega}\left(e\right) \ , \nonumber \\
\frac{dM}{dt} &=& n
+\omega_3B_N'\left(\frac{r_0}{a}\right)^\kappa\frac{q}{m}
g_{\kappa,M}\left(e\right)\cos i \ .
\eeqa

Here, $g_{\kappa,\omega}$, $g_{\kappa,\Omega}$, and $g_{\kappa, M}$ denote
the eccentricity functions of the angle-like variables, provided also, in 
Appendix A. We notice, that the secular components of the angle-like variables are stemming
from the terms independent of the solar wind velocity -- opposite to the
secular terms of the action-like variables in \RI{Equation} \equ{secgauss}. \\

We now directly compare a numerical integration of the original \RI{set
of Equations} \equ{EOM} for $\beta=0$, $m_1=0$ with the secular system \RI{defined by
Equations} \equ{secgauss},\equ{secgauss2}, using $\varphi=0$, $\vartheta=0$, 
$\kappa=2$, and $T=22yrs$. We choose for the normal magnetic field component:

\beq{toyMF}
b_N(t)=1+\cos\left(2\pi t/T\right) \ .
\eeq

The choice is motivated to allow $B_N$ to be of non-zero average, with the
additional property to be periodic in the magnetic solar cycle. \RI{The 
constant in Equation \equ{toyMF} is used to demonstrate its effect
on the secular evolution of the orbital elements, i.e. drift. The periodic
term ensures that the normal component varies with the same period as the
remaining components of the magnetic field, i.e. $B_R$, $B_T$ in
Equation \equ{dy3}. While the existence of a non-zero normal component of the 
magnetic field has already been observed, e.g. in \citet{Zur1997} one may question 
the validity of Equation \equ{toyMF}. Does a persistent and significant non-zero 
value for $B_N$ exist? Indeed there are indications that a north-south asymmetry in
the solar magnetic field causes a persistent average cone-shaped
asymmetry of the heliospheric magnetic field, which is called the "bashful ballerina"
\citep[][]{HilMur2006, MurVir2012}. We estimate the magnitude of $B_N$ from 
measurements of the Ulysses spacecraft during its polar orbit around the Sun. We make use
of the analysis of the data provided in \citet{For2002}, where the authors
investigate deviations from the standard Parker model in terms of the meridional 
angle $\delta_B$. In this setting the magnitude of $B_N$ may be obtained from 
$\sin\delta_B=B_N/|\mathbf B|$ (see appendix) for given $\delta_B$ and $|\mathbf B|$.} 

\begin{figure}
\begin{center}
\includegraphics[width=.4\linewidth]{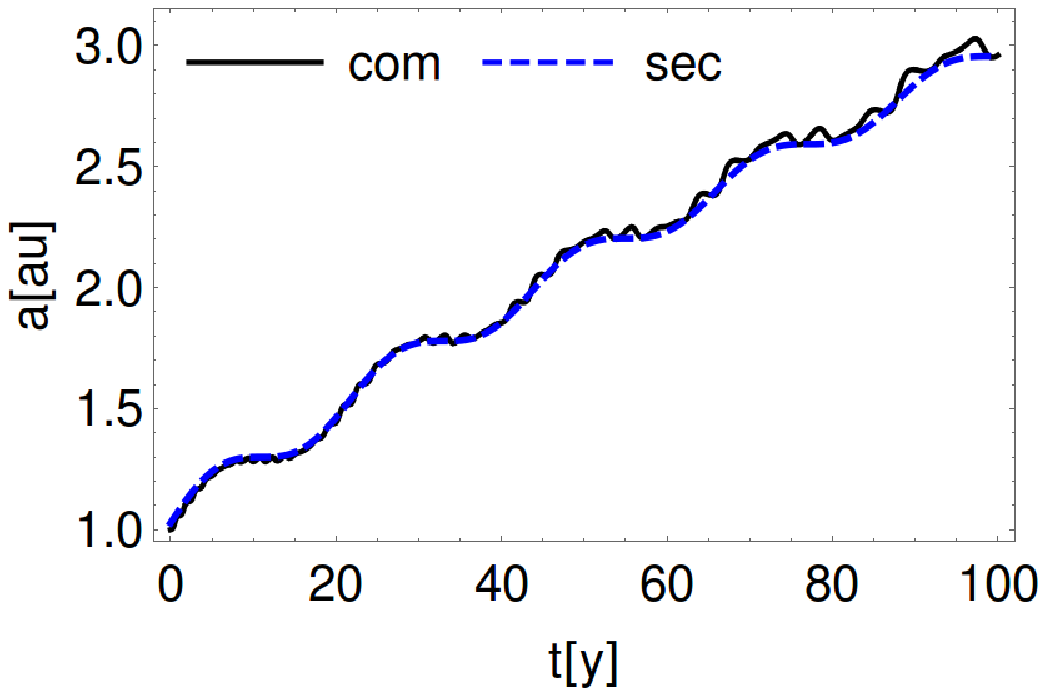}
\includegraphics[width=.4\linewidth]{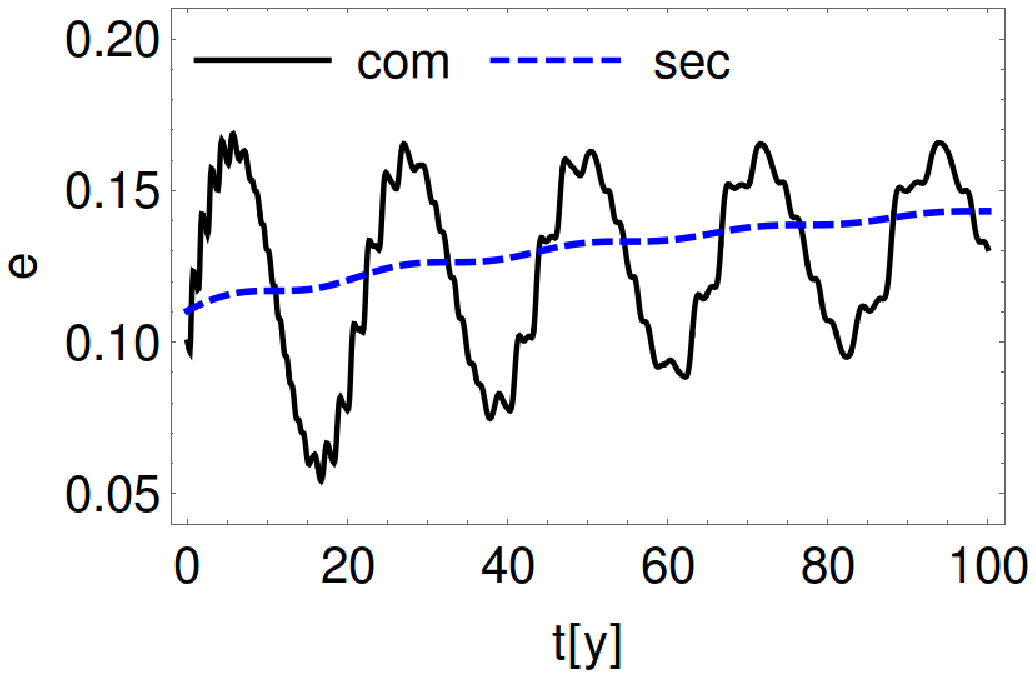}
\includegraphics[width=.4\linewidth]{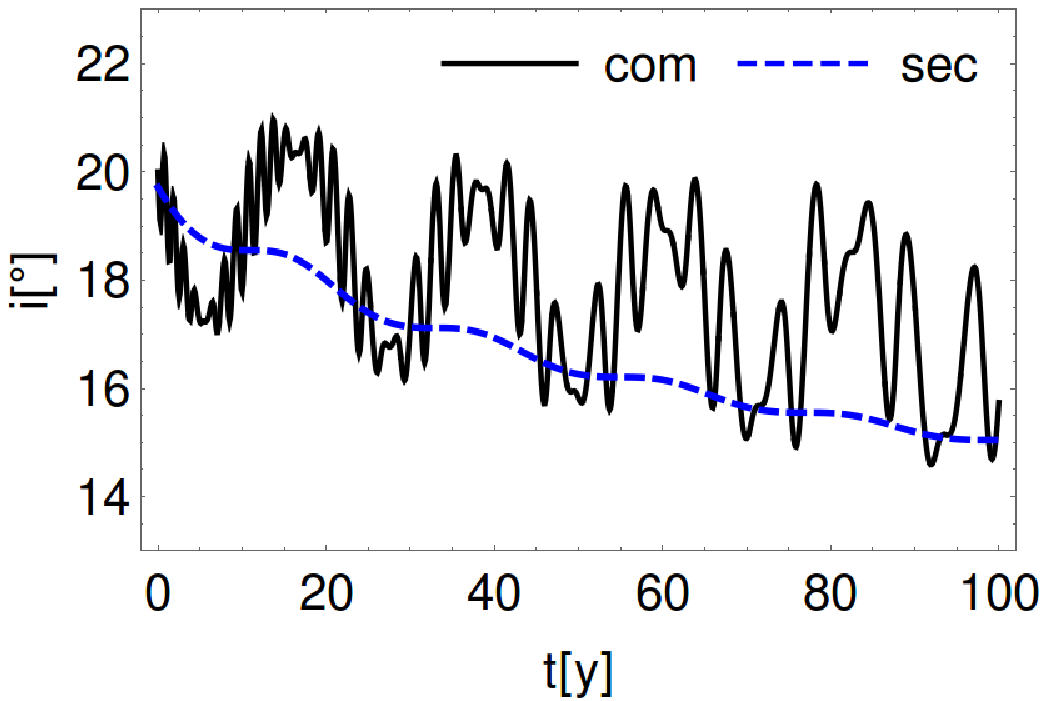}
\includegraphics[width=.4\linewidth]{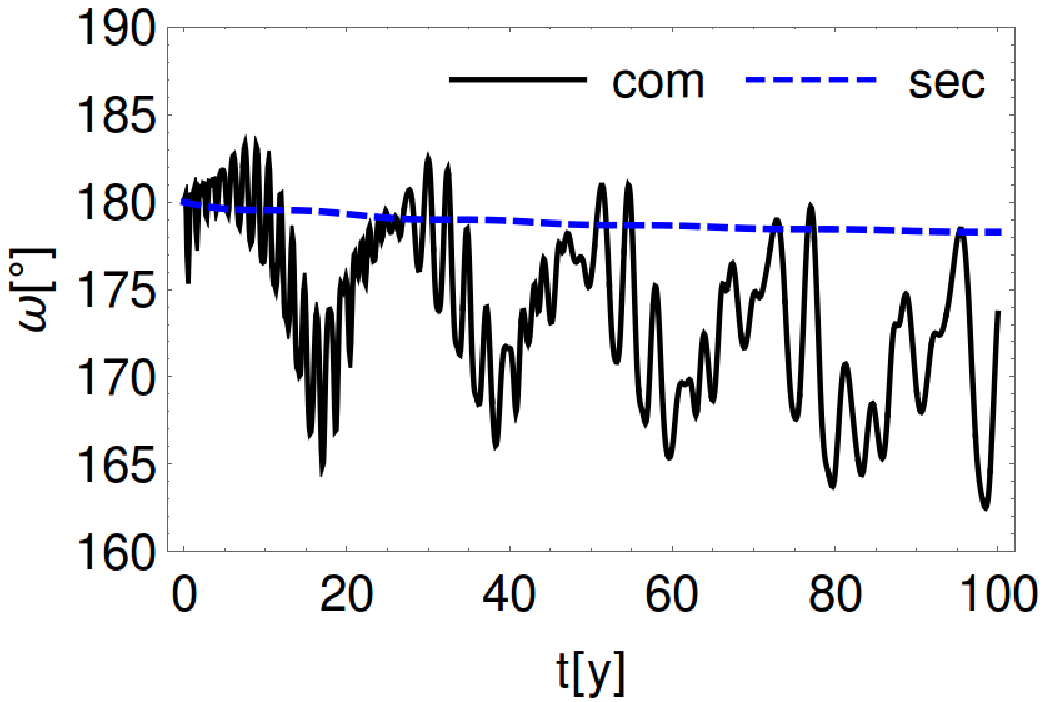}
\includegraphics[width=.4\linewidth]{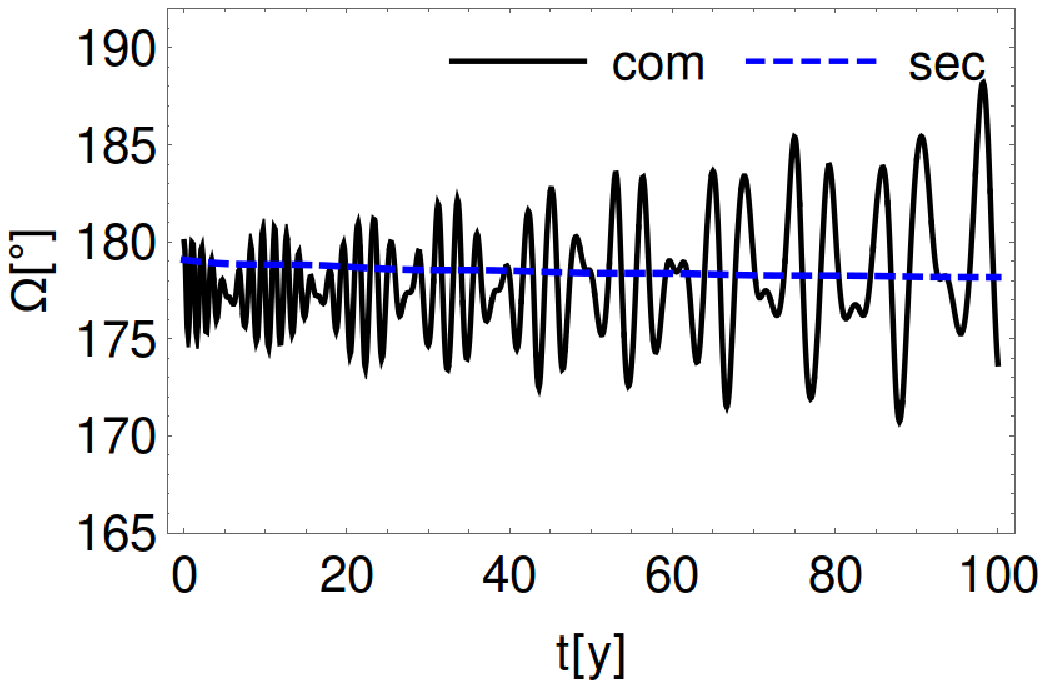}
\includegraphics[width=.4\linewidth]{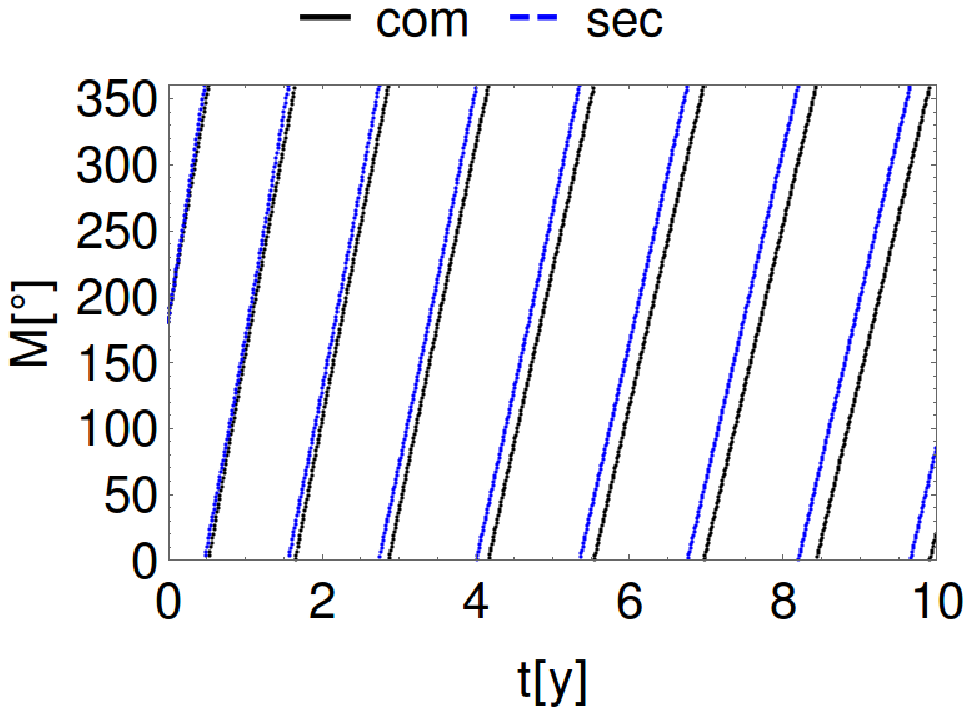}
\end{center}
\caption{Comparison of \RI{Equation} \equ{EOM} ('com'plete, black) and \RI{Equations}  
\equ{secgauss},\equ{secgauss2} ('sec'ular, dashed blue) dynamics based 
on \RI{Equation} \equ{toyMF}. From upper left to lower right: semi-major axis $a$, 
eccentricity $e$, inclination $i$, argument of perihelion $\omega$, longitude of 
ascending node $\Omega$, mean anomaly $M$.}
\label{toy}
\end{figure}

In Figure~\ref{toy} we show the comparison between the complete and simplified
system. We find that \RI{Equations} \equ{secgauss},\equ{secgauss2} well
describe the secular dynamics of the complete system \RI{defined by Equation}
\equ{EOM} on secular time scales. The orbit corresponds to the motion of
charged particles with radius $R=1\mu m$, density $\rho=2g/cm^3$, with the
surface electric potential of \RI{$U=+5V$}. For the simulation we use
\RI{$B_{R0}=B_{T0}=3\times10^{-9} \ T$ and $B_{N0}=0.5\times10^{-9} \ T$} at
$r_0=1au$. We choose the slow solar wind speed $u_{sw}=400km/s$, and set the
magnetic axis to $\omega_3=\cos(7.25{}^o)$.  Initial conditions are $a(0)=1$,
$e(0)=0.1$, $i(0)=20{}^o$, $\omega(0)=\Omega(0)=M(0)=180{}^o$, respectively.

\subsection{Near Hamiltonian form}

Let $L=\sqrt{\mu a}$, $G=L\sqrt{1-e^2}$, $H=G\cos i$, $l=M$, $g=\omega$, and $h=\Omega$ 
be the proper action-angle variables to our problem (usually referred to as  
Delaunay variables). The derivatives of these variables with respect to time are: 

\beqa{HAM}
\frac{dL}{dt}&=&\frac{\mu}{2L}\frac{da}{dt} \ , 
\quad \frac{dl}{dt}=\frac{dM}{dt} \ , \nonumber \\
\frac{dG}{dt}&=&\frac{\mu G}{2L^2}\frac{da}{dt} - \frac{L\sqrt{L^2-G^2}}{G}\frac{de}{dt} \ ,
\quad \frac{dg}{dt}=\frac{d\omega}{dt} \ , \nonumber  \\
\frac{dH}{dt}&=&-\frac{\mu H}{2L^2}\frac{da}{dt} + \frac{HL\sqrt{L^2-G^2}}{G}\frac{de}{dt}
-\sqrt{G^2-H^2}\frac{di}{dt} \ ,
\quad \frac{dh}{dt}=\frac{d\Omega}{dt} \ .
\eeqa

We remark that, due to the velocity dependent terms in \RI{Equation} \equ{EOM},
the system \RI{of Equations} \equ{HAM} can not be derived from a potential for the velocity
dependent terms that are associated to solar wind and Poynting-Robertson drag,
as well as the magnetic field. However, \RI{Equations} \equ{HAM} are still of the
special form:

\beqa{HAM2}
&&\frac{dL}{dt}=-\frac{d\mathfrak H}{dl} + {\mathfrak f}_L \ , \
\frac{dG}{dt}=-\frac{d\mathfrak H}{dg} + {\mathfrak f}_G \ , \
\frac{dH}{dt}=-\frac{d\mathfrak H}{dh} + {\mathfrak f}_H\ , \nonumber \\
&&\frac{dl}{dt}=\frac{d\mathfrak H}{dL} + {\mathfrak f}_l\ , \
\frac{dg}{dt}=\frac{d\mathfrak H}{dG} + {\mathfrak f}_g\ , \
\frac{dh}{dt}=\frac{d\mathfrak H}{dH} + {\mathfrak f}_h \ , \
\eeqa

\noindent where ${\mathfrak H}$ denotes the Hamiltonian part, and ${\mathfrak
f}_L$, ${\mathfrak f}_G$, ${\mathfrak f}_H$ as well as ${\mathfrak f}_l$,
${\mathfrak f}_g$, ${\mathfrak f}_h$ denote the non-conservative parts of
\RI{Equation} \equ{EOM}, only. For $\beta=0$, $q/m=0$ the underlying dynamical
system is integrable, thus \RI{Equation} \equ{HAM2} -- and therefore also
\RI{Equation} \equ{EOM} -- is a nearly conservative, weakly dissipative
dynamical system \citep[see][]{CL2012, LC2013}. The total time derivative of
the Hamiltonian is given by:

\beq{dHdt}
\Phi=
\frac{\partial {\mathfrak H}}{\partial L}\frac{dL}{dt} +
\frac{\partial {\mathfrak H}}{\partial G}\frac{dG}{dt} +
\frac{\partial {\mathfrak H}}{\partial H}\frac{dH}{dt} +
\frac{\partial {\mathfrak H}}{\partial l}\frac{dl}{dt} +
\frac{\partial {\mathfrak H}}{\partial g}\frac{dg}{dt} +
\frac{\partial {\mathfrak H}}{\partial h}\frac{dh}{dt} +
\frac{d {\mathfrak H}}{dt}\ .
\eeq

Taking for $\mathfrak H$ only the terms in \RI{Equation} \equ{EOM} that correspond to the
two-body problem, but taking for the time derivatives of the Delaunay variables,
\RI{Equation} \equ{HAM2} we find:

\beq{dHdtC}
\Phi = \frac{\mu^2}{2a^2}\frac{da}{dt} \ ,
\eeq

with $da/dt$ taken from \RI{Equation} \equ{secgauss}.

\subsection{Balance of forces}

We are interested in the long-term evolution of charged particles subject to Lorentz force,
solar radiation pressure, solar wind, and Poynting-Robertson drag forces. As we
see in Figure~\ref{toy}, the magnetic field of the solar system leads to an increase of 
mean semi-major axis $a$ in time $t$. This effect supersedes the solar wind and 
Poynting-Robertson drag that would otherwise lead to a decrease of $a$. The secular 
components of these forces in Delaunay variables \citep[][]{JanLem2001} reduce to: 

\beq{SWPR}
\frac{dL}{dt} = -\gamma
\frac{n\left(1+\frac{3}{2}e^2\right)}{\left(1-e^2\right)^{3/2}} \ , \quad
\frac{dG}{dt} = -\gamma \ n \ , \quad
\frac{dH}{dt} = -\gamma \ n \cos i \ ,
\eeq

while the averaged effect of the drag forces on the angular variables $l$, $g$,
$h$ turns out to vanish. We notice, that the common proportionality factor
$\gamma$, different from the definition of \citet{JanLem2001}, is simply given
by:

\beq{dy5}
\gamma = \frac{\beta\mu}{c}\left(1+\frac{\eta}{Q}\right) \ .
\eeq

By making use of \RI{Equation} \equ{HAM} we are able to express \RI{Equation}
\equ{SWPR} in terms of $da/dt$, $de/dt$, $di/dt$, respectively. In this
section, we focus on solutions where the inward drift in the semi-major axis
due to Poynting-Robertson and solar wind drag is balanced with the outward
drift due to Lorentz-force. This is the case if the first \RI{of Equation}
\equ{secgauss} equals to $-da/dt$ stemming from the first of \RI{Equation}
\equ{SWPR}. By proper arrangement of terms we obtain the condition:
 
\beq{formula}
\frac{q}{m}=
\frac{\beta}{c}
\frac{\left(Q+\eta\right)}{Q}
\frac{n^3a^{\kappa+2}G_\kappa\left(e\right)}{r_0^{\kappa}\cos\left(i\right)}
\left(B_N'u_{sw}\omega_3\right)^{-1} \ ,
\eeq

where the functions $G_\kappa=G_\kappa(e)$ are given in Appendix A. We notice
that the above relation is the condition on the physical parameters $q/m$,
$\beta$, $Q$ of a micrometer-sized dust particle to be secularly stable for the
mean of the orbital elements $n$, $a$, $e$, and $i$, in a solar wind and
magnetic field environment parametrized by $\eta$, $B_N'$, $u_{sw}$, and
$\omega_3$.  Let us denote the surface potential $U$ associated to
\RI{Equation} \equ{formula} by the equilibrium surface potential from know on.\\



\subsection{Additional gravitational perturbations}

In the previous sections we have disregarded the influence of the planets by
setting $m_1=0$. This assumption can only be true in regimes of motion within
our solar system, where dust grains are not in mean motion resonance with the
other planets. For a discussion of the role of such planetary mean motion
resonances see \citet{Wei1993, Sic1993, Der1994, BeaFer1994} or \citet{Lio1997,
Koc2008, Pas2009, LhoCel2015}. We can expect a quite different dynamical
picture for charged dust grains in a resonant lock with the planets as compared
to the purely dissipative case. The time of temporary capture and the locations
of the resonant regimes of motion in the solar system will be strongly effected
by the actual charge of the dust grains, and moreover, by the actual
interplanetary magnetic field. This topic deserves further investigations, but
is beyond the scope of this study.

\section{Numerical simulations \& parameter study}
\label{parker}

We aim in this Section, first, to confirm our analytical results by
means of direct numerical integration of \RI{Equation} \equ{EOM}, and second,
we aim to investigate the long term dynamics of dust particles with
different exponents $\kappa$ that determine $B_N$ in \RI{Equation} \equ{Bex}.
We study parameters and initial conditions that are close to
the balanced solutions of \RI{Equation} \equ{EOM}, in particular those solutions
that remain close to their initial values on secular times. Furthermore,
we perform a parameter study valid for secularly stable, electrically
charged dust grains. \\

In Section 2 we used \RI{Equation} \equ{EOM} with $B_N$ \RI{given by Equation} \equ{Bex} and 
$b_N$ \RI{taken from Equation} \equ{toyMF} with $\kappa=2$. Here, we use
$\kappa=1,2,3$ together with the physical parameters that we summarize in Table
~\ref{tab:1}. Astronomical and physical constants are taken from \citet{AstCon,
TDP}. Ranges for various parameters are derived on the basis of proposed values
in literature: \citet{BLS1979, Gus1994, GruEtAl1994}, and  \citet{Koc2006}.
Typical optical properties and densities, that are consistent with observations
yield $\beta\simeq0.2/R$ with $R$ given in $[\mu m]$ \citep[][]{BeaFer1994},
and $\beta\simeq7.6\times10^{-4}A/m$ with $A/m$ given in $[\rm m^2/kg]$
\citep[][]{Koc2006}.\\

\begin{table}
\label{tab:1}
$$
\begin{array}{|l|c|l|c|}
\hline
\# & value & unit & ref. \\
\hline
\beta & 0\ldots0.5 & - & \text{\cite{Gus1994}} \\
q/m & -0.5\ldots0.5 & C kg^{-1} & \text{v.i.} \\
Q & 1\ldots 2 & - & \text{\cite{Koc2006}} \\
R & 0.5\ldots10 & \mu m & \text{\cite{GruEtAl1994}} \\
\rho & 0.5\ldots2 & g cm^{-3} & \text{\cite{GruEtAl1994}} \\
U & -10\ldots10 & V & \text{\cite{GruEtAl1994}} \\
\hline
\RI{B_{R0}}, \RI{B_{T0}} & 3.0\times10^{-9} & T & \text{\cite{Koc2006}} \\
\RI{B_{N0}} & \RI{0.5\times10^{-9}} & \RI{T} & \RI{\text{v.i.}} \\
c & 299792458 & km s^{-1} & \text{\cite{AstCon}} \\
\varepsilon_0 & 8.854 187\times10^{-12} & F/m & \text{\cite{TDP}} \\
Gm_1 & 1.266 865\times10^{17} & m^3s^{-2} & \text{\cite{AstCon}} \\
\eta & 1/3 & - & \text{\cite{Koc2006}} \\
\mu & 1.327 124\times10^{20} & m^3s^{-2} & \text{\cite{AstCon}} \\
S \ (1au) & 1.3608 & kW/m^2 & \text{\cite{BLS1979}} \\
u_{sw} & 400 & km s^{-1} & \text{\cite{GruEtAl1994}} \\
(\omega_1,\omega_2,\omega_3) & (0.035,0.121,0.992) & - & \text{v.i.} \\
\hline\hline
\end{array}
$$
\caption{Physical parameters. The charge is obtained from $q=4\pi \varepsilon_0 U R$, 
mass from $m=4/3\pi\rho R^3$, cross section from $A=\pi R^2$. \RI{Normal component $B_{N0}$
estimated from $B_{N0}=|B_{R0}|\sin\delta_B$ with meridional angle $\delta_B\simeq10^o$
\citep[][]{For2002}.} 
The direction of the magnetic $z$-axis with respect to the inertial plane is given by
$\omega_1=\cos\ascnode\sin\iota$,
$\omega_2=\sin\ascnode\sin\iota$,
$\omega_3=\cos\iota$ with $\ascnode=73.67{}^o$ and $\iota=7.25{}^o$, taken
from \citet{Eps1917}.}
\end{table}

\begin{figure}
\centering
\includegraphics[width=0.45\linewidth]{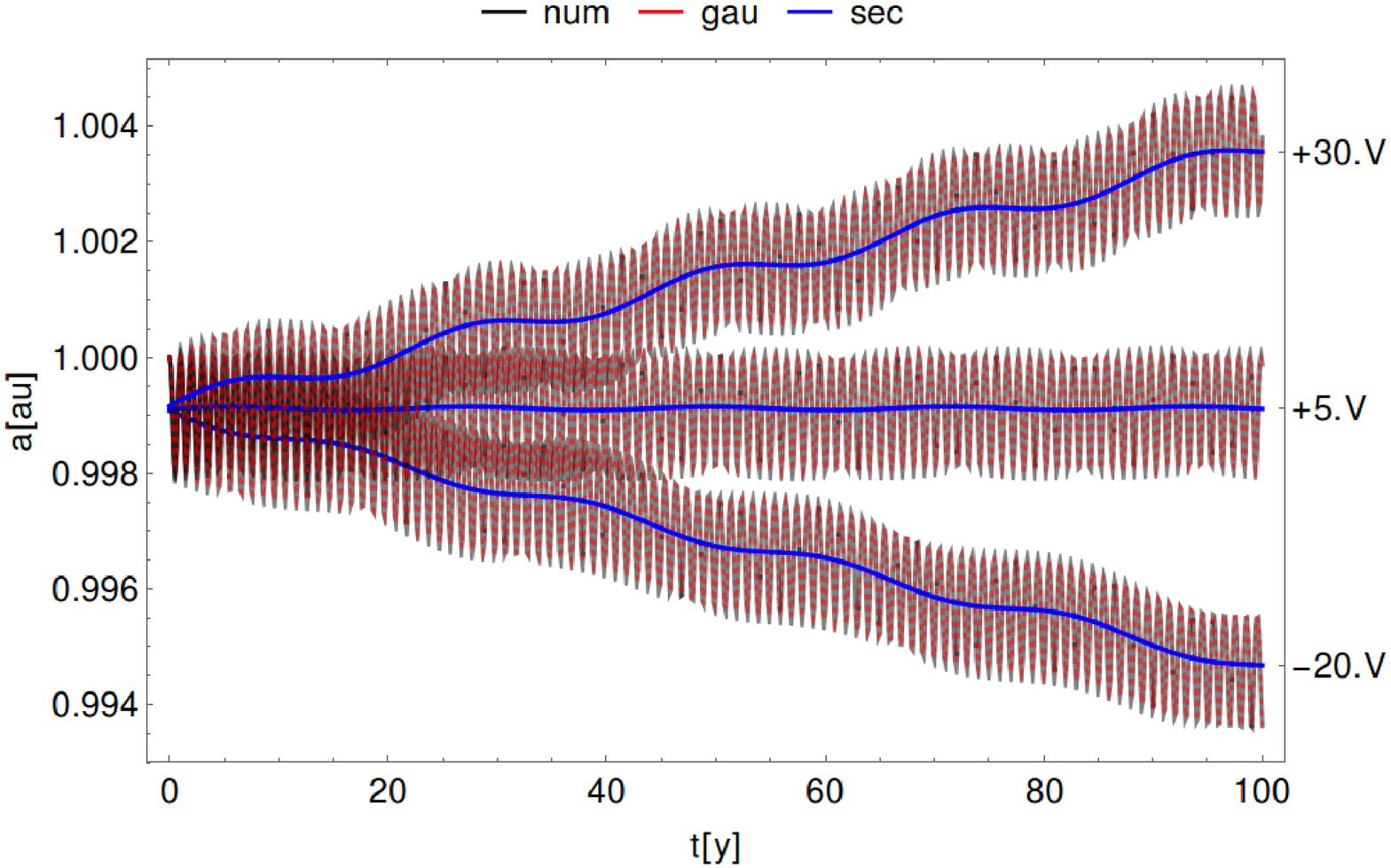}
\includegraphics[width=0.45\linewidth]{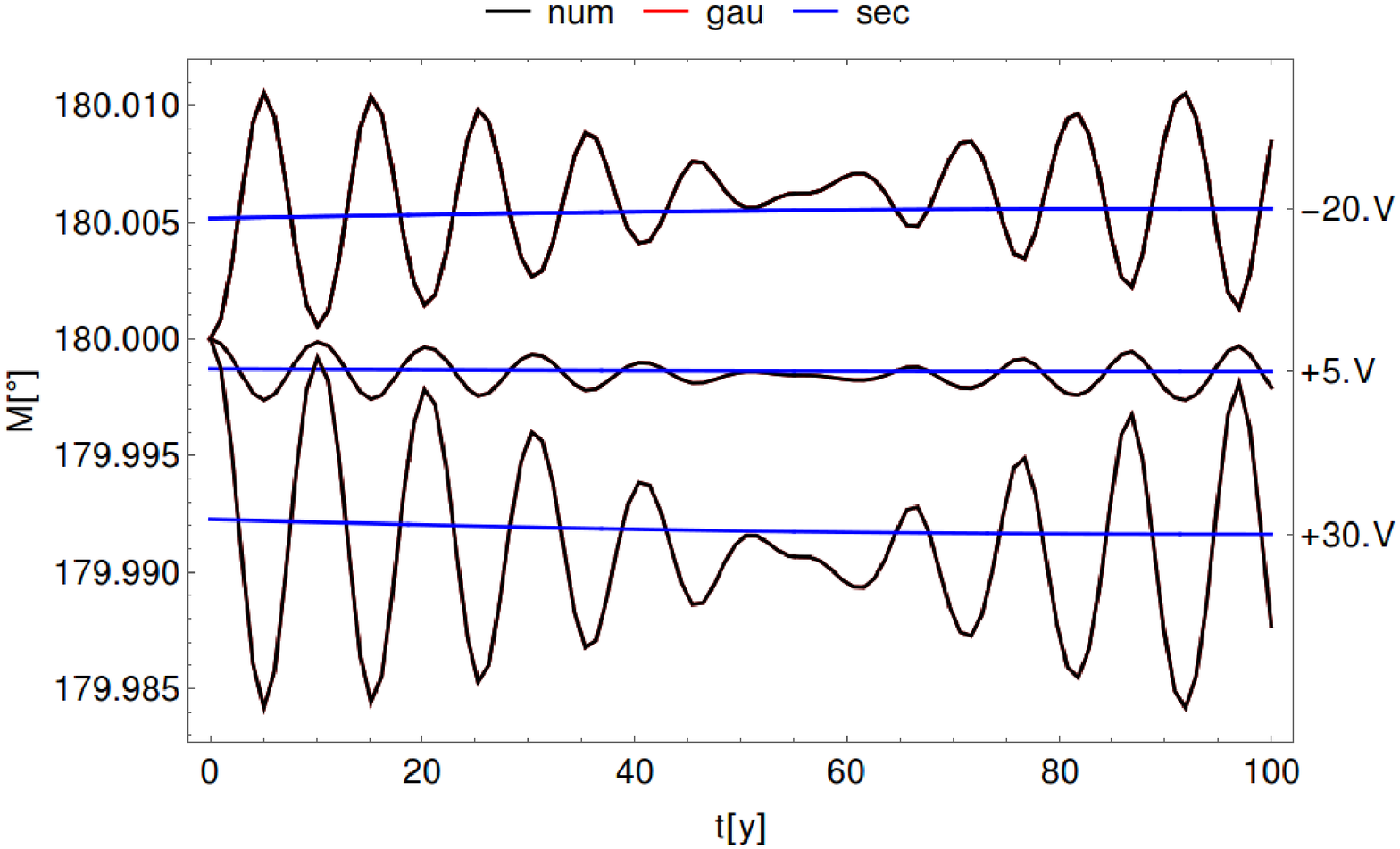}
\includegraphics[width=0.45\linewidth]{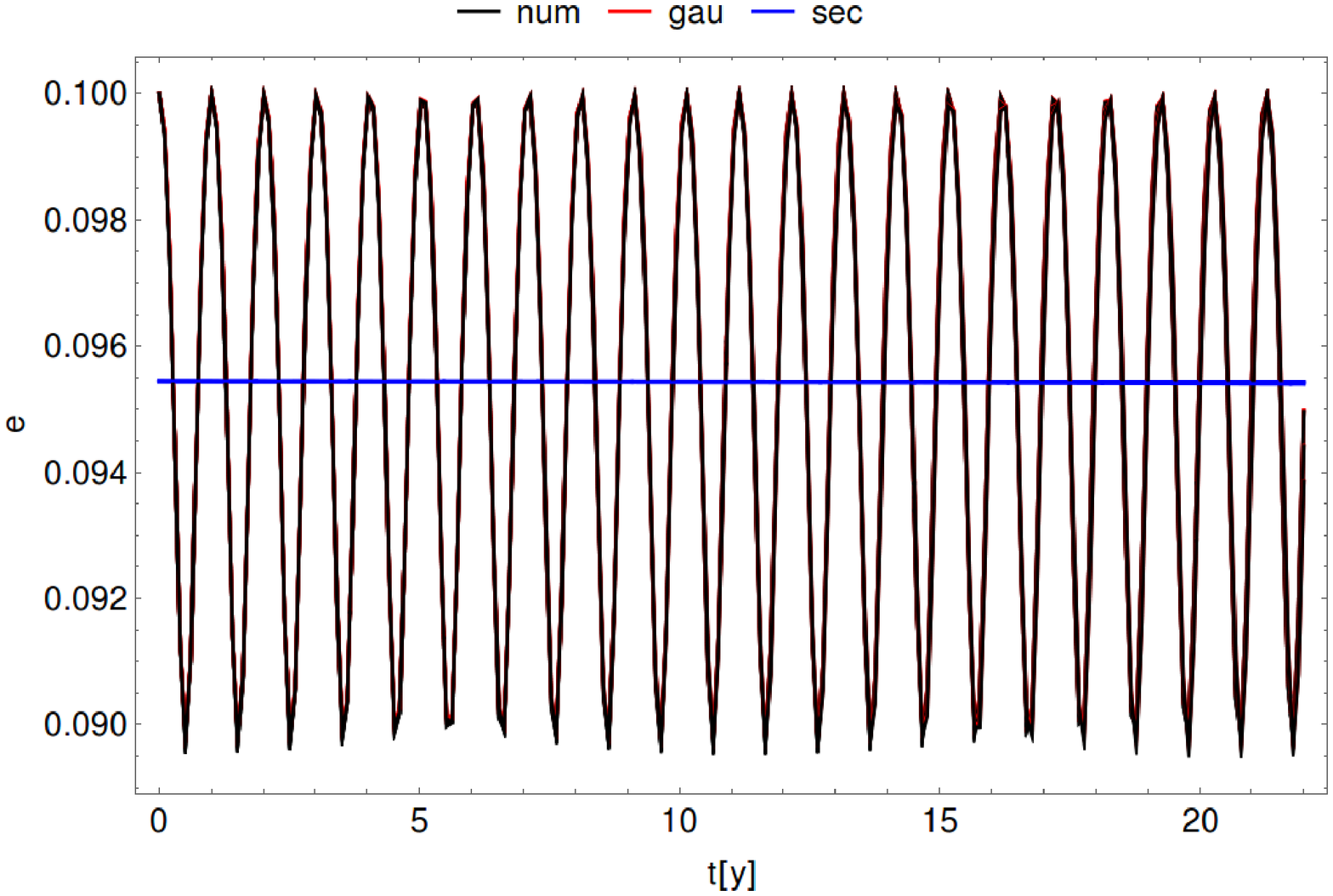}
\includegraphics[width=0.45\linewidth]{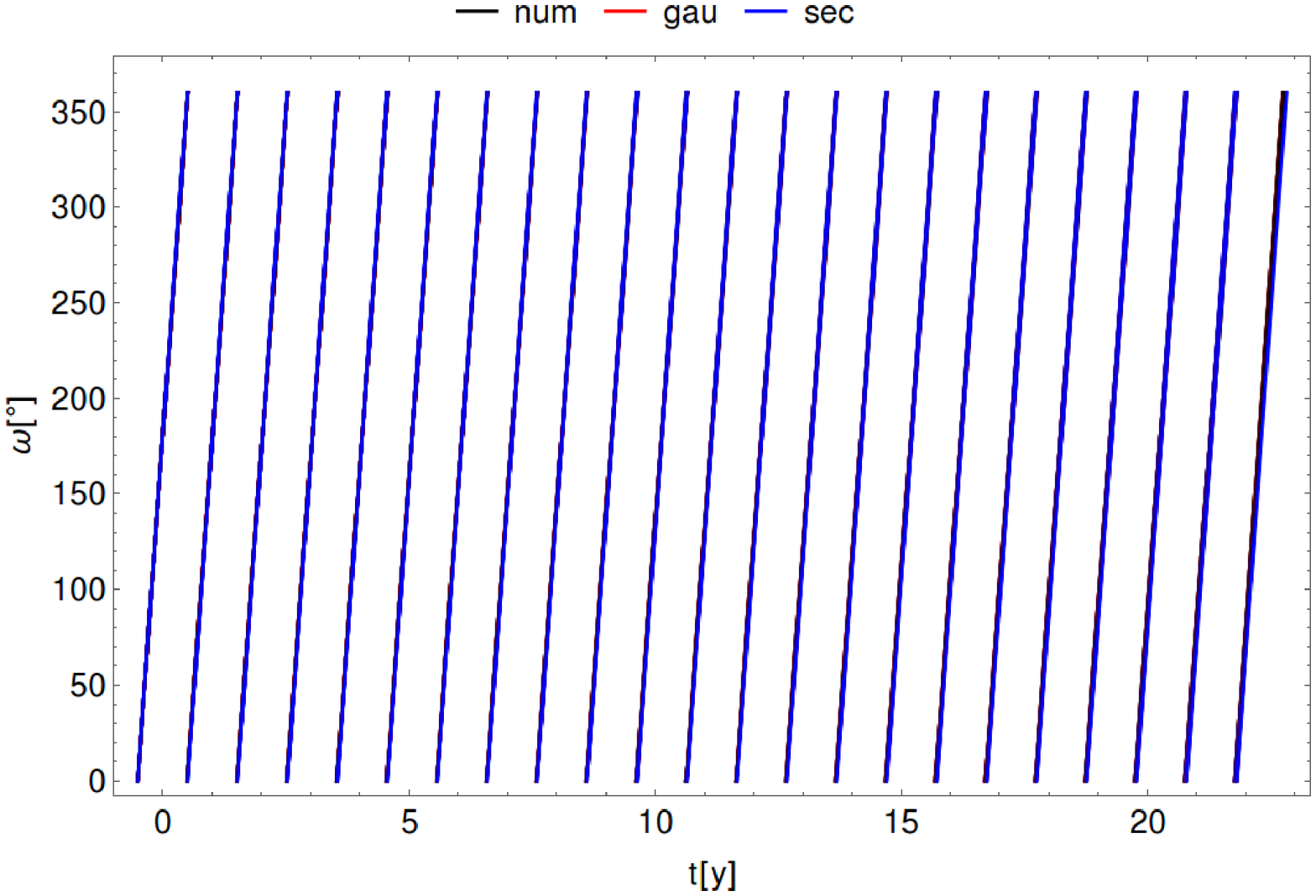}
\includegraphics[width=0.45\linewidth]{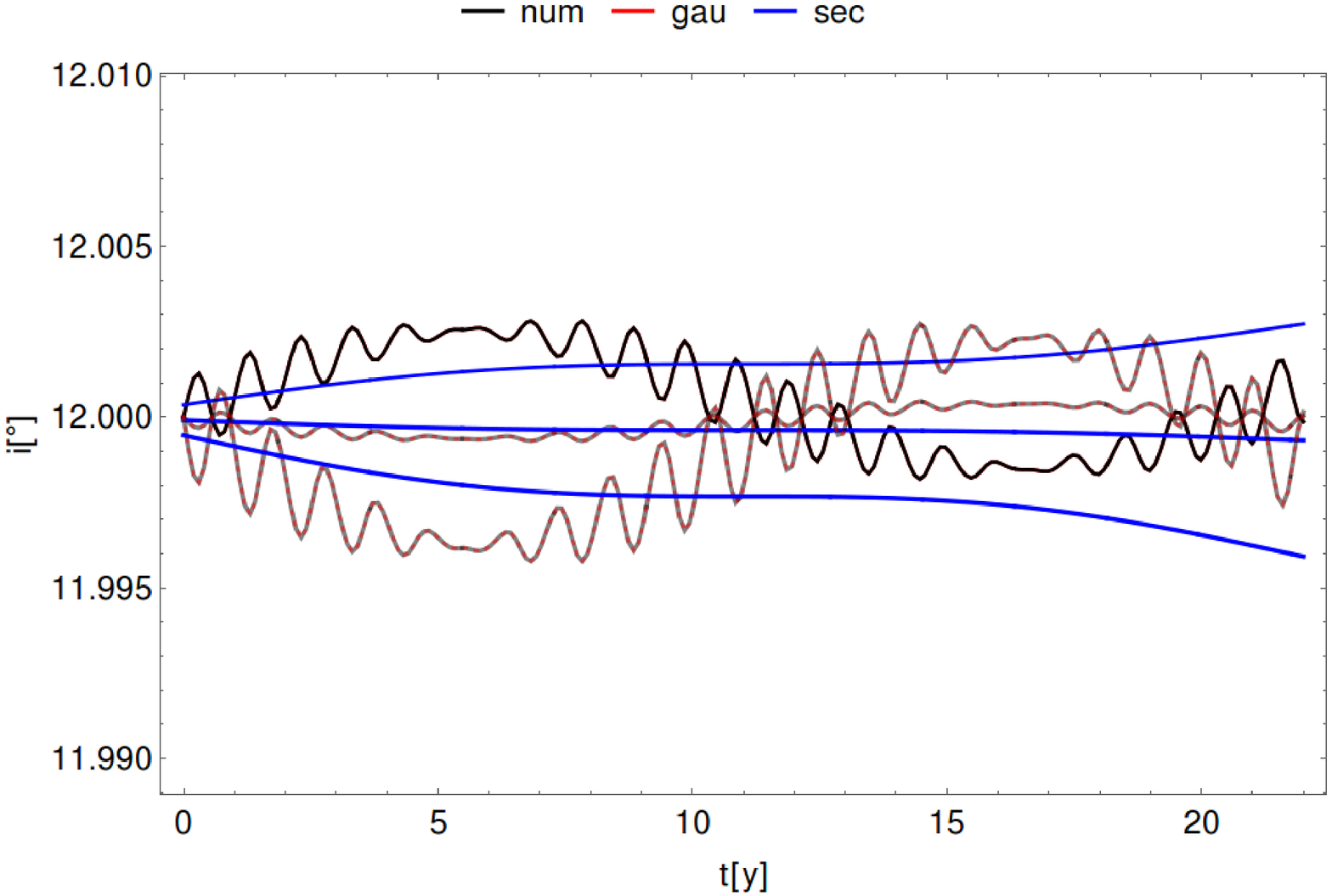}
\includegraphics[width=0.45\linewidth]{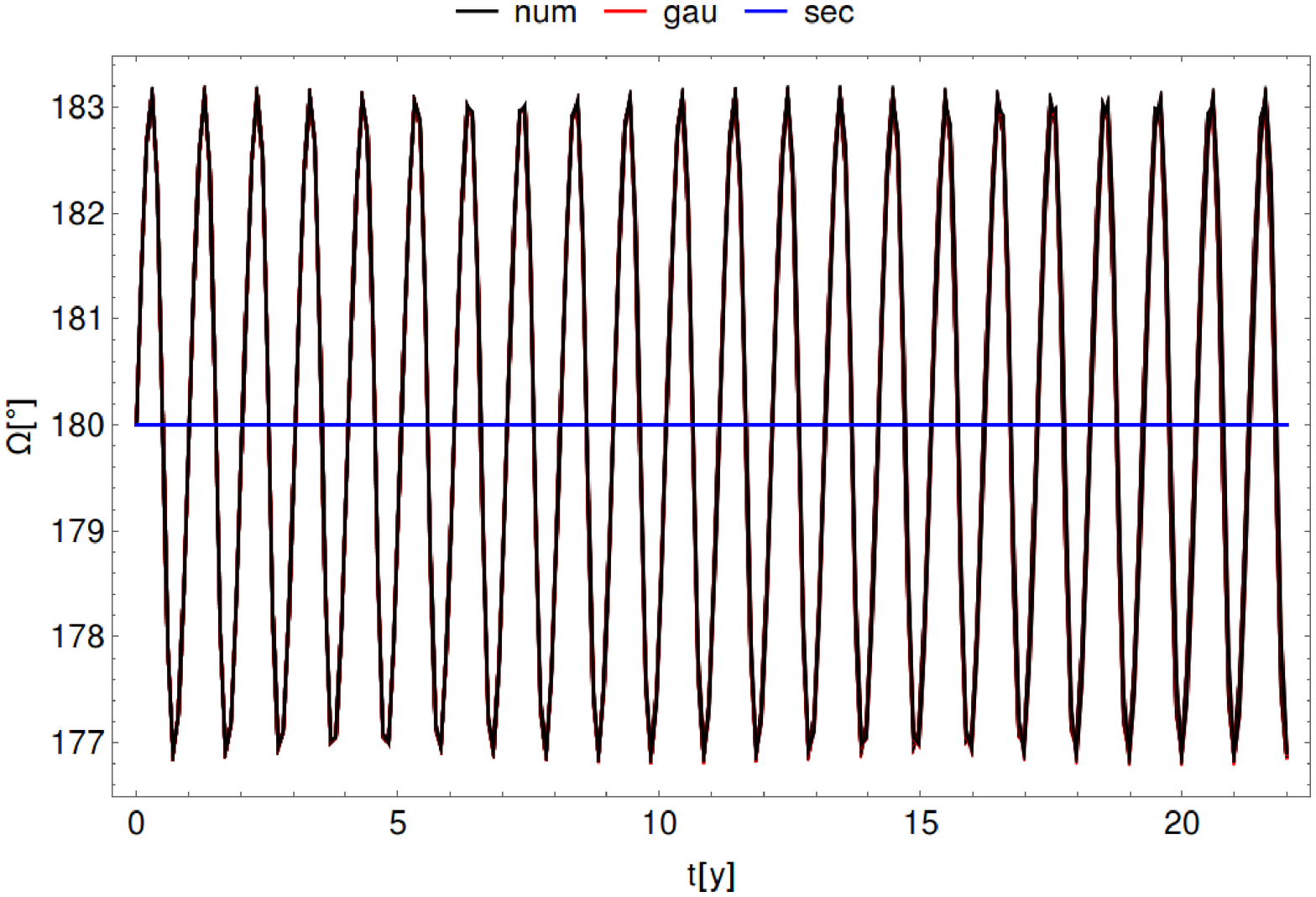}
\caption{Orbital dynamics of charged dust grains in the vicinity of the
equilibrium surface potential \RI{(5V)}. Numerical solutions based on
\RI{Equation} \equ{EOM} (black, {\it num}), \RI{Equation} \equ{gauss} (red,
{\it gau}), and \RI{Equations} \equ{secgauss},\equ{secgauss2} (blue, {\it
sec}), respectively. Small frame ticks on the right indicate surface potential
in Volts.}
\label{num1}
\end{figure}

In Figure~\ref{num1} we show typical orbits close to a secular equilibrium.
\RI{We set $\kappa=1$,  $U=5V$, $Q=1$, $\rho=2g/cm^3$, and $a(0)=1au$,
$e(0)=0.1$, $i(0)=12{}^o$, together with initial angles set to
$\omega(0)=\Omega(0)=M(0)=180{}^o$. From Equation \equ{formula} we find
$q/m\simeq2.1\times10^{-5}$ that corresponds to $R\simeq55.5\mu m$ and
$\beta\simeq0.005$. For this values} we find a secularly stable orbit on time
scales longer than several solar cycles. This solution therefore stays close to
its initial semi-major axis, as expected. A higher charge leads to a positive
drift, a lower charge to a negative drift in the semi-major axis. For
completeness, we also show time series of the remaining orbital elements: the
period of the mean anomaly $M$ is nearly constant as we can see in the
top-right panel of Figure~\ref{num1}. The eccentricity $e$ of the orbit
oscillates around a mean value close to its initial condition, that is only
slightly affected by changing $q/m$, while the argument of perihelion $\omega$
rotates.  The inclination $i$ of the charged particle stays nearly constant.
Finally, the longitude of the ascending node oscillates around its initial
condition with very similar amplitudes for different $q/m$ ratios. We like to
point out that the essential orbital dynamics is reproduced in all three
dynamical models. However, the drift in inclination is overestimated in the
purely secular model as one can see in the lower-left panel of
Figure~\ref{num1}. \\

\begin{figure}
\centering
\includegraphics[width=0.65\linewidth]{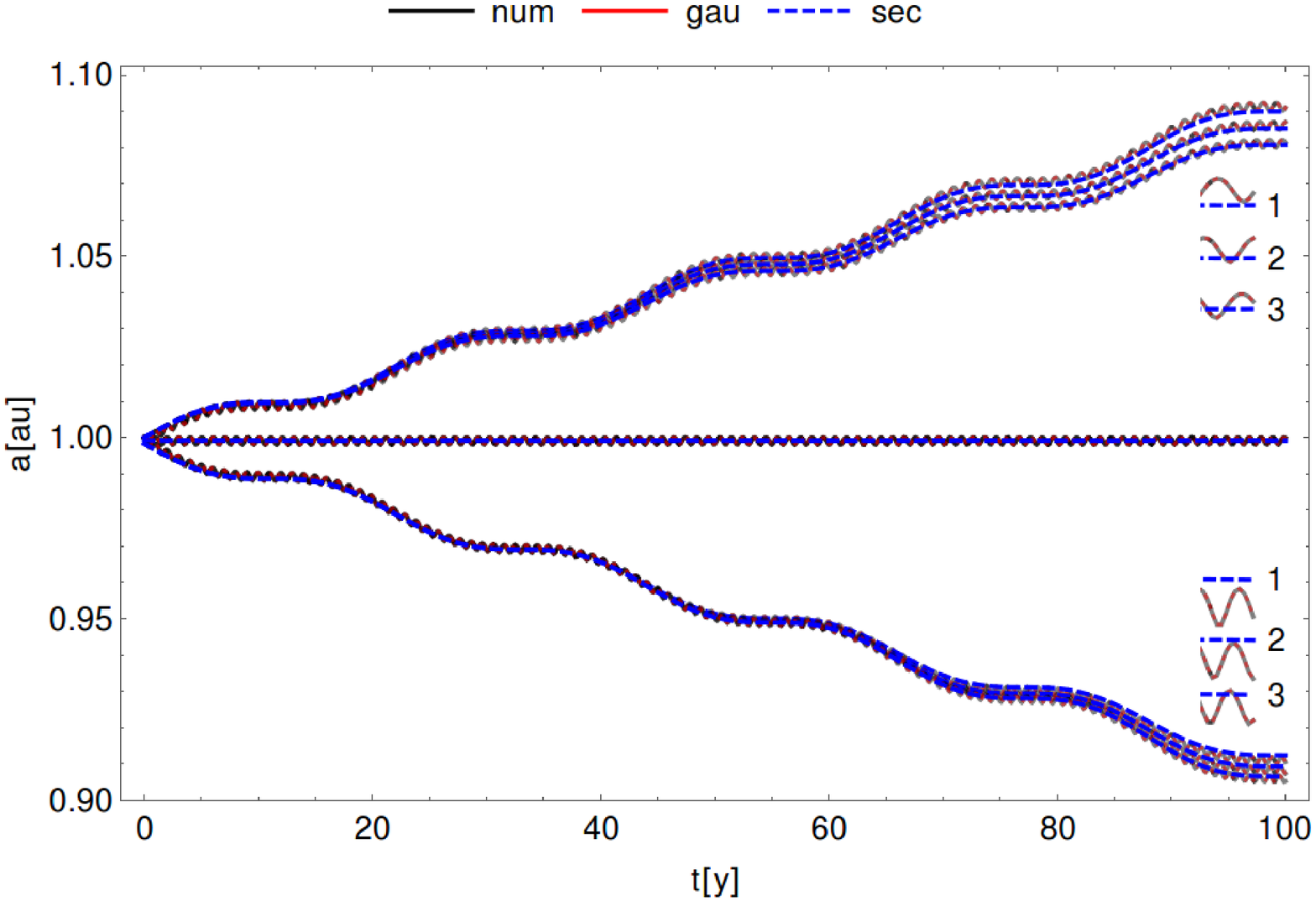}
\caption{Orbital dynamics of charged dust grains in the vicinity of the equilibrium
surface potential \RI{(5V)}. Numerical solutions based on \RI{Equation} \equ{EOM} 
(black, {\it num}), \RI{Equation} \equ{gauss} (red, {\it gau}), and \RI{Equations} 
\equ{secgauss},\equ{secgauss2} (blue, {\it sec}), respectively.
\RI{Ticks inside the figure indicate $\kappa$ in \RI{Equation} \equ{Bex}.}}
\label{num2}
\end{figure}

If we repeat our study for $\kappa=2,3$ we find the same kind of dynamical
behaviour of the particles close to the balanced solutions, see
Figure~\ref{num2}: for $q/m$ that leads to secularly stable motions the drift
in semi-major axis remains close to zero. For larger deviations
\RI{($\pm500V$)} from the value $q/m$ obtained from \RI{Equation} \equ{formula}
\RI{corresponding to $5V$} the situation changes as follows: for the lower
value of $q/m$ the negative drift in semi-major axis $a$ becomes stronger for
increasing $\kappa$. For larger values of $q/m$ the positive drift becomes
weaker. The other orbital elements are only slightly perturbed for increasing
$\kappa$, while keeping the structural form of the solution (not shown here).
\\

\begin{figure}
\centering
\includegraphics[width=0.65\linewidth]{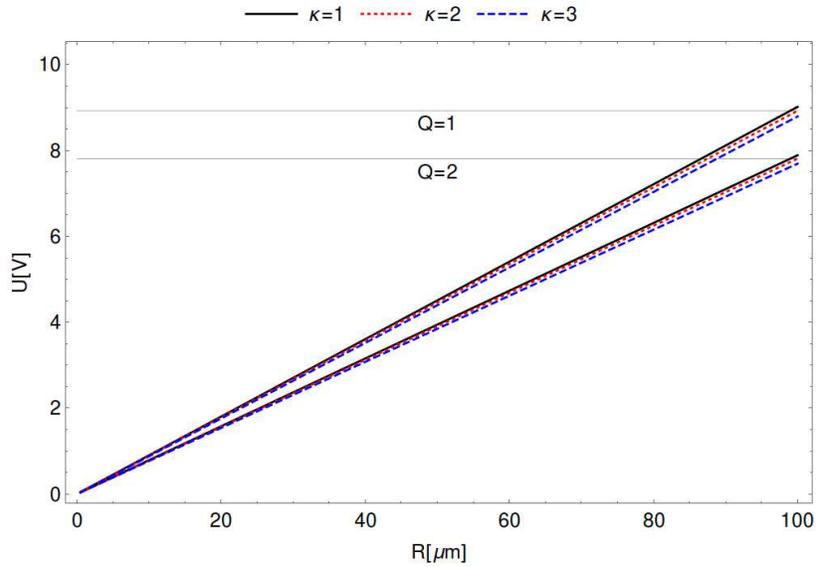}
\caption{Equilibrium surface potential $U$ (in Volts) of charged
dust grains at $1au$ for different radii $R$ (in $\mu m$) and $\kappa$.}
\label{stud1}
\end{figure}

\begin{figure}
\centering
\includegraphics[width=0.65\linewidth]{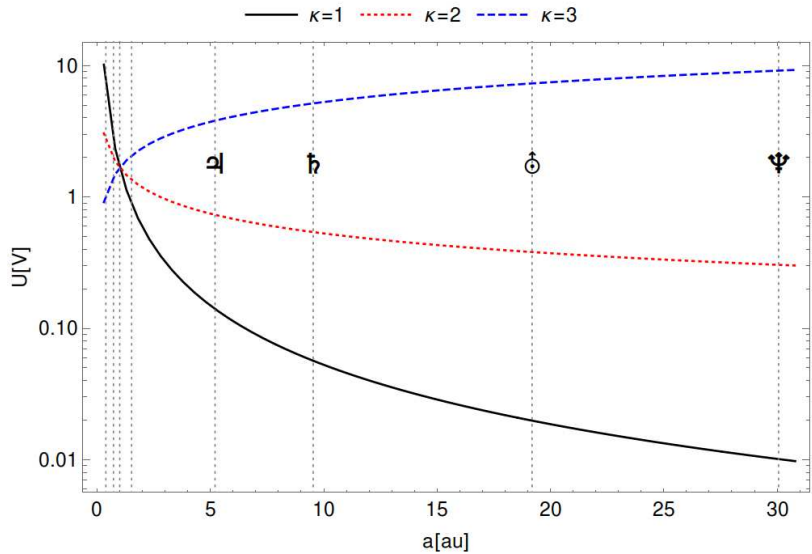}
\caption{Parameter study of the equilibrium surface potential $U$ (in Volts) of charged dust 
grains with fixed radius \RI{$R=100\mu m$} for different exponents $\kappa$ in \RI{Equation} 
\equ{Bex}. Dotted lines indicate the semi-major axes of the planets.}
\label{stud2}
\end{figure}

We also perform a parameter study in $R$, $\rho$, and $Q$ to check for the dependency
of the surface potential $U$ that leads to secularly stable motions, on some typical 
properties of micro-meter sized dust grains. In Figure~\ref{stud1} we show how
the equilibrium voltage for a dust grain located at $1au$ mainly depends on the radius $R$ 
of the grain. It depends to about $10-15\%$ on the quality factor $Q$, while a dependency 
on the density $\rho$ would not be visible. We find that $U$ also depends on the exponent
$\kappa$, i.e. the parametrization of the interplanetary magnetic field. We also perform a 
second parameter study in varying distance $a$ from the Sun for fixed dust particle 
characteristics. In Figure~\ref{stud2} we clearly see the strong
dependency of the equilibrium surface potential $U$ on the interplanetary magnetic field
model, i.e. the radial dependence parametrized by the exponent $\kappa$.

\section{Summary \& Conclusions}
\label{sum}

The orbital stability of charged dust grains in our solar system is strongly
affected by various non-gravitational forces, i.e. by the solar
wind and the Poynting-Robertson drag forces, as well as the Lorentz force from the
interplanetary magnetic field. We therefore investigate the combination of these
effects and their influence on the long term dynamics of charged dust grains. Major 
discoveries from our study on the role of the magnetic field in the micrometer-sized 
particle dynamics are:

\begin{enumerate}

\item The normal component of the magnetic field strongly affects the long-term
stability of charged dust particles, leading to secular positive or negative
drift in the semi-major axis depending on the actual charge over mass ratio. The 
normal component of the magnetic field is usually not included in standard models of the
interplanetary magnetic field \citep{Parker1958, weber1967}. More realistic
magnetic field models that include a normal component will improve our 
understanding of long-term dust dynamics.

\item There are special values of charge over mass ratios that balances out the
solar wind and Poynting-Robertson drag forces with the Lorentz force at given
distance from the Sun. This $q/m$ ratios lead to secularly stable motions, and
depend on the amplitude of the normal component of the interplanetary magnetic field, 
the orbital shapes, and the physical properties of the charged particles. The 
measurement of electric charge of dust grains during \RI{future} interplanetary space missions 
will allow to test our predictions of secularly stable particle orbits for specific 
$q/m$ ratios.

\end{enumerate}

Moreover, we provide a qualitative description of the dynamics of charged dust
grains on the basis of averaging theory: the secular drift in the semi-major
axis of charged dust grains turns out to be proportional to the strength of the
normal component of the interplanetary magnetic field, the mean solar wind
speed, and proportional to the inverse arbitrary exponent $\kappa$ applied to
the distance from the Sun. The direction of the drift (outwards or inwards from
the initial distance) strongly depends on the value and sign of the charge over
mass ratio.  The secular evolution in orbital eccentricity $e$, and inclination
$i$ is of the order of $1/a$ times the drift in the semi-major axis. The drift
in $a$ and $e$ turns out to be more effective for smaller inclinations $i$. On
the contrary, the drift in inclination $i$ itself becomes less effective in
smaller $i$. The ratio $q/m$, that yields secularly stable motions, is
proportional to $\beta$ that depends on the radius, the density, and optical
properties of the dust grain.  It increases along with the distance from the
Sun for $\kappa=3$ and decreases with $\kappa=1,2$. The charge over mass ratio
$q/m$ that leads to secularly stable motions turns out to be slightly larger
for smaller efficiency factors $Q$ and smaller $\kappa$. If these kinds of
charged dust grains exist in large amounts then the interplanetary medium may
contain a significant amount of charged particles of same parity because there
is an asymmetry in the balance of forces: solar wind and Poynting-Robertson
drag can only add negative $da/dt$, while Lorentz force can contribute with
$\pm da/dt$. However, only positively charged dust grains may counteract the
negative drift if the average of $B_N$ is positive (and vice versa).
Therefore, one can expect a larger amount of charged particles of same parity
if the mean of the normal field component is non-zero.

Our study is the first of a series of studies on the interplay of the
interplanetary magnetic field and charged dust grains within the solar
system. We still neglect the gravitational influence of the major
bodies and use a simplified model of the interplanetary magnetic
field (i.e. we omit interplanetary magnetic sectors of opposite polarity,
and local perturbations of the magnetic field). In addition,
applying the assumptions of a radial magnetic field leads to an impossible
magnetic monopole in the center of a Parker spiral.
However, we would like to stress that \RI{Equation} \equ{formula} holds true for more
generic normal magnetic field components $B_N$. It would therefore
be very interesting to clarify if a non-zero average component of the 
interplanetary magnetic field (or z-component of the solar magnetic axis) 
exists, at least for sufficient long enough periods of time to trigger
secular motions in the orbital dynamics of electrically charged
dust grains.

Typical applications of our work are: the dust environment of the solar system
in general, i.e. dust that is released by asteroid collisions or by means of
cometary activity. The dust environment in the vicinity of the moons and the
planets, in particular the Lagrange points of the system. \RI{Dust experiments
have become part of important interplanetary space missions: the Ulysses
spacecraft that was the first to study the Sun from pole to pole. The space
probe measurements include the solar wind, charged particles, neutral gases and
small particles from the local interstellar space.  The Galileo dust detector
on board of the Galileo spacecraft was intended to measure dust grains over a
wide range of masses in interplanetary space and in the Jovian system.
Measurements provide physical and dynamical properties as functions of the
distances to the Sun, to Jupiter, and the Jovian satellites.  The cosmic dust
analyzer on board of the Cassini spacecraft measured the chemical composition
of dust during its cruise to Saturn, investigated the Io dust streams during
its Jupiter flyby, mapped the size distribution of ring material, and analyzed
gravitationally bound ejecta particles in the vicinity of the icy satellites.
Future space probes} should be able to measure: the dust kinematics (velocity
vectors), the dust properties (size, weight, density, optical properties,
temperature, chemical composition, and charge), as well as the normal component
of the magnetic field environment.  The measurements will not only provide
important new insights into the structure of the interplanetary magnetic field,
it will also allow to study the physical properties of dust grains that may
have strong implications on coagulation and planet formation. Last but not
least, it will allow to test the hypothesis of an electrically charged solar
system, in terms of secularly stable, electrically charged dust particle
orbits.


\begin{appendices}

\section*{Appendix A}

The secular system \RI{defined by Equations} \equ{secgauss},\equ{secgauss2} comprises the 
eccentricity functions $g_{\kappa,j}$, valid up to $O(e^5)$, that are given in Table~\ref{AA}. 
The eccentricity functions $G_\kappa(e)$ in \RI{Equation} \equ{formula} for $\kappa\in\{1,2,3\}$,
expanded up to $O(e^5)$ are:

\beqa{dy7}
G_{1}(e)&=&1+3e^2+33e^4/8 , \nonumber \\
G_{2}(e)&=&1+2e^2+ 9e^4/8 , \nonumber \\
G_{3}(e)&=&1+e^2/2-9e^4/8 .
\eeqa

A Taylor series expansion in $e$ of the right hand sides of \RI{Equation} \equ{gauss}, i.e.
$c_{\mathbf k}$, $s_{\mathbf k}$, may be obtained by one of the authors.

\begin{table}
$$
\begin{array}{||l|ccc||}
\hline\hline
\kappa & 1 & 2 & 3 \\
\hline
g_{\kappa,a} &  2 & 2+2 e^2+2 e^4 & 2+5 e^2+8 e^4 \\
g_{\kappa,e} & -e/2+e^3/8 & e/2+e^3/8 
& 3 e/2+3 e^3/2 \\
g_{\kappa,i} & -1/2-e^2/4-3 e^4/16 
& -1/2-e^2/4-3e^4/16 & -1/2-e^2/2-e^4/2 \\
g_{\kappa,\omega} &  -1/2 & -1-5 e^2/8 & -3/2-9 e^2/4 \\
g_{\kappa,\Omega} & -1/2 & -1/2-e^2/4-3 e^4/16 
& -1/2-3e^2/4-15 e^4/16 \\
g_{\kappa,a} & -1+e^2/2 & -1/2+e^2/8 & 0 \\
\hline\hline
\end{array}
$$
\caption{Eccentricity functions $g_{\kappa,j}$ with $j\in\{a,e,i,\omega,\Omega,M$\} for
$\kappa=1,2,3$, respectively.}
\label{AA}
\end{table}

\end{appendices}

\section*{Acknowledgements}

\RI{We thank M. Bentley for providing material on measurements of physical dust
properties. We also thank an anonymous reviewer who greatly helped us to improve
a previous version of the manuscript.}


\bibliography{biblio}

\begin{thebibliography}{}
\expandafter\ifx\csname natexlab\endcsname\relax\def\natexlab#1{#1}\fi

\bibitem[{{Beauge} \& {Ferraz-Mello}(1994)}]{BeaFer1994}
{Beauge}, C., \& {Ferraz-Mello}, S. 1994, Icarus, 110, 239

\bibitem[{{Burns} {et~al.}(1979){Burns}, {Lamy}, \& {Soter}}]{BLS1979}
{Burns}, J.~A., {Lamy}, P.~L., \& {Soter}, S. 1979, Icarus, 40, 1

\bibitem[{{Celletti} \& {Lhotka}(2012)}]{CL2012}
{Celletti}, A., \& {Lhotka}, C. 2012, Regular and Chaotic Dynamics, 17, 273

\bibitem[{{Consolmagno}(1979)}]{Con1979}
{Consolmagno}, G.~J. 1979, \icarus, 38, 398

\bibitem[{{Dermott} {et~al.}(1994){Dermott}, {Jayaraman}, {Xu}, {Gustafson}, \&
  {Liou}}]{Der1994}
{Dermott}, S.~F., {Jayaraman}, S., {Xu}, Y.~L., {Gustafson}, B.~{\AA}.~S., \&
  {Liou}, J.~C. 1994, \nat, 369, 719

\bibitem[{{Dvorak} \& {Lhotka}(2013)}]{mybook}
{Dvorak}, R., \& {Lhotka}, C. 2013, {Celestial Dynamics: Chaoticity and
  Dynamics of Celestial Systems} (Wiley-VCH)

\bibitem[{{Epstein}(1917)}]{Eps1917}
{Epstein}, T. 1917, Astronomische Nachrichten, 204, 351

\bibitem[{{Fahr} {et~al.}(1981){Fahr}, {Ripken}, \& {Lay}}]{FahEtAl1981}
{Fahr}, H.~J., {Ripken}, H.~W., \& {Lay}, G. 1981, \aap, 102, 359

\bibitem[{{Fahr} {et~al.}(1995){Fahr}, {Scherer}, \&
  {Banaszkiewicz}}]{FahEtAl1995}
{Fahr}, H.~J., {Scherer}, K., \& {Banaszkiewicz}, M. 1995, \planss, 43, 301

\bibitem[{{Fitzpatrick}(1970)}]{Fitzpatrick}
{Fitzpatrick}, P.~M. 1970, {Principles of celestial mechanics} (Academic Press,
  New York)

\bibitem[{{Forsyth} {et~al.}(1996){Forsyth}, {Balogh}, {Horbury}, {Erdoes},
  {Smith}, \& {Burton}}]{For1996}
{Forsyth}, R.~J., {Balogh}, A., {Horbury}, T.~S., {et~al.} 1996, Astronomy \&
  Astrophysics, 316, 287

\bibitem[{{Forsyth} {et~al.}(2002){Forsyth}, {Balogh}, \& {Smith}}]{For2002}
{Forsyth}, R.~J., {Balogh}, A., \& {Smith}, E.~J. 2002, Journal of Geophysical
  Research (Space Physics), 107, 1405

\bibitem[{{Gr{\"u}n} {et~al.}(1994){Gr{\"u}n}, {Gustafson}, {Mann}, {Baguhl},
  {Morfill}, {Staubach}, {Taylor}, \& {Zook}}]{GruEtAl1994}
{Gr{\"u}n}, E., {Gustafson}, B., {Mann}, I., {et~al.} 1994, A\&A, 286, 915

\bibitem[{{Gustafson}(1994)}]{Gus1994}
{Gustafson}, B.~A.~S. 1994, Annual Review of Earth and Planetary Sciences, 22,
  553

\bibitem[{{Hiltula} \& {Mursula}(2006)}]{HilMur2006}
{Hiltula}, T., \& {Mursula}, K. 2006, Geophysical Research Letters, 33, L03105

\bibitem[{{Jancart} \& {Lemaitre}(2001)}]{JanLem2001}
{Jancart}, S., \& {Lemaitre}, A. 2001, Celestial Mechanics and Dynamical
  Astronomy, 81, 75

\bibitem[{{Kla{\v c}ka}(2013)}]{Kla2013}
{Kla{\v c}ka}, J. 2013, \mnras, 436, 2785

\bibitem[{{Kocifaj} \& {Kla{\v c}ka}(2004)}]{KocKla2004}
{Kocifaj}, M., \& {Kla{\v c}ka}, J. 2004, \planss, 52, 839

\bibitem[{{Kocifaj} \& {Kla{\v c}ka}(2008)}]{Koc2008}
---. 2008, \aap, 483, 311

\bibitem[{{Kocifaj} {et~al.}(2006){Kocifaj}, {Kla{\v c}ka}, \&
  {Horvath}}]{Koc2006}
{Kocifaj}, M., {Kla{\v c}ka}, J., \& {Horvath}, H. 2006, MNRAS, 370, 1876

\bibitem[{{Krivov} {et~al.}(1998{\natexlab{a}}){Krivov}, {Kimura}, \&
  {Mann}}]{KriEtAl1998}
{Krivov}, A., {Kimura}, H., \& {Mann}, I. 1998{\natexlab{a}}, \icarus, 134, 311

\bibitem[{{Krivov} {et~al.}(1998{\natexlab{b}}){Krivov}, {Mann}, \&
  {Kimura}}]{KriManKim1998}
{Krivov}, A., {Mann}, I., \& {Kimura}, H. 1998{\natexlab{b}}, Earth, Planets,
  and Space, 50, 551

\bibitem[{{Lhotka} \& {Celletti}(2013)}]{LC2013}
{Lhotka}, C., \& {Celletti}, A. 2013, International Journal of Bifurcation and
  Chaos, 23, 50036

\bibitem[{{Lhotka} \& {Celletti}(2015)}]{LhoCel2015}
---. 2015, Icarus, 250, 249

\bibitem[{{Liou} \& {Zook}(1997)}]{Lio1997}
{Liou}, J.-C., \& {Zook}, H.~A. 1997, \icarus, 128, 354

\bibitem[{{Luzum} {et~al.}(2011){Luzum}, {Capitaine}, {Fienga}, {Folkner},
  {Fukushima}, {Hilton}, {Hohenkerk}, {Krasinsky}, {Petit}, {Pitjeva},
  {Soffel}, \& {Wallace}}]{AstCon}
{Luzum}, B., {Capitaine}, N., {Fienga}, A., {et~al.} 2011, Celestial Mechanics
  and Dynamical Astronomy, 110, 293

\bibitem[{{Mann} {et~al.}(2006){Mann}, {K{\"o}hler}, {Kimura}, {Cechowski}, \&
  {Minato}}]{ManEtAl2006}
{Mann}, I., {K{\"o}hler}, M., {Kimura}, H., {Cechowski}, A., \& {Minato}, T.
  2006, \aapr, 13, 159

\bibitem[{{Mann} {et~al.}(2014){Mann}, {Meyer-Vernet}, \&
  {Czechowski}}]{ManEtAl2014}
{Mann}, I., {Meyer-Vernet}, N., \& {Czechowski}, A. 2014, \physrep, 536, 1

\bibitem[{{Mann} {et~al.}(2007){Mann}, {Murad}, \&
  {Czechowski}}]{ManMurCze2007}
{Mann}, I., {Murad}, E., \& {Czechowski}, A. 2007, \planss, 55, 1000

\bibitem[{Meyer-Vernet(2007)}]{Mey2007}
Meyer-Vernet, N. 2007, Basics of the Solar Wind, Cambridge Atmospheric and
  Space Science Series (Cambridge University Press)

\bibitem[{{Morfill} \& {Gr{\"u}n}(1979{\natexlab{a}})}]{MorGru1979b}
{Morfill}, G.~E., \& {Gr{\"u}n}, E. 1979{\natexlab{a}}, PSS, 27, 1283

\bibitem[{{Morfill} \& {Gr{\"u}n}(1979{\natexlab{b}})}]{MorGru1979a}
---. 1979{\natexlab{b}}, PSS, 27, 1269

\bibitem[{{Mukai} \& {Giese}(1984)}]{MukGie1984}
{Mukai}, T., \& {Giese}, R.~H. 1984, \aap, 131, 355

\bibitem[{{Mursula} \& {Virtanen}(2012)}]{MurVir2012}
{Mursula}, K., \& {Virtanen}, I.~I. 2012, Journal of Geophysical Research, 117,
  A08104

\bibitem[{{Parker}(1958)}]{Parker1958}
{Parker}, E.~N. 1958, ApJ, 128, 664

\bibitem[{{P{\'a}stor} {et~al.}(2009){P{\'a}stor}, {Kla{\v c}ka}, \&
  {K{\'o}mar}}]{Pas2009}
{P{\'a}stor}, P., {Kla{\v c}ka}, J., \& {K{\'o}mar}, L. 2009, Celestial
  Mechanics and Dynamical Astronomy, 103, 343

\bibitem[{{Ragot} \& {Kahler}(2003)}]{RagKah2003}
{Ragot}, B.~R., \& {Kahler}, S.~W. 2003, \apj, 594, 1049

\bibitem[{{Sicardy} {et~al.}(1993){Sicardy}, {Beauge}, {Ferraz-Mello},
  {Lazzaro}, \& {Roques}}]{Sic1993}
{Sicardy}, B., {Beauge}, C., {Ferraz-Mello}, S., {Lazzaro}, D., \& {Roques}, F.
  1993, Celestial Mechanics and Dynamical Astronomy, 57, 373

\bibitem[{{St\"oecker}(2014)}]{TDP}
{St\"oecker}, H. 2014, {Taschenbuch der Physik} (Edition Hari Deutsch)

\bibitem[{{Wallis} \& {Hassan}(1985)}]{WalHas1985}
{Wallis}, M.~K., \& {Hassan}, M.~H.~A. 1985, \aap, 151, 435

\bibitem[{{Weber} \& {Davis}(1967)}]{weber1967}
{Weber}, E.~J., \& {Davis}, Jr., L. 1967, ApJ, 148, 217

\bibitem[{{Weidenschilling} \& {Jackson}(1993)}]{Wei1993}
{Weidenschilling}, S.~J., \& {Jackson}, A.~A. 1993, \icarus, 104, 244

\bibitem[{{Zurbuchen} {et~al.}(1997){Zurbuchen}, {Schwadron}, \&
  {Fisk}}]{Zur1997}
{Zurbuchen}, T.~H., {Schwadron}, N.~A., \& {Fisk}, L.~A. 1997, J. Geo. Res.,
  102, 24175

\end{thebibliography}

\bibliographystyle{apj}

\end{document}